\documentclass[sigconf]{acmart}
\usepackage{subfigure}
\usepackage{geometry}
\usepackage{soul}
\usepackage{url}
\usepackage{epstopdf}
\usepackage{multirow}
\usepackage{multicol}
\usepackage{booktabs}
\usepackage{bbm}
\usepackage{graphicx}
\usepackage{float}
\usepackage{color}
\usepackage{eqparbox}
\newcommand\ie{\textit{i.e., }}
\newcommand\eg{\textit{e.g., }}
\newcommand\modelname{HGPF}
\newcommand{\tabincell}[2]{\begin{tabular}{@{}#1@{}}#2\end{tabular}} 
\usepackage[linesnumbered, ruled]{algorithm2e}
\usepackage{latexsym}
\usepackage{bm}
\usepackage{amsmath}
\usepackage{multirow}
\usepackage{amsfonts}
\SetKwRepeat{Do}{do}{while}%

\AtBeginDocument{%
  \providecommand\BibTeX{{%
    \normalfont B\kern-0.5em{\scshape i\kern-0.25em b}\kern-0.8em\TeX}}}

\copyrightyear{2023} 
\acmYear{2023} 
\setcopyright{acmlicensed}\acmConference[WWW '23]{Proceedings of the ACM Web Conference 2023}{May 1--5, 2023}{Austin, TX, USA}
\acmBooktitle{Proceedings of the ACM Web Conference 2023 (WWW '23), May 1--5, 2023, Austin, TX, USA}
\acmPrice{15.00}
\acmDOI{10.1145/3543507.3583282}
\acmISBN{978-1-4503-9416-1/23/04}

\begin{document}

\title{A Post-Training Framework for Improving Heterogeneous Graph Neural Networks}
\author{Cheng Yang}
\email{yangcheng@bupt.edu.cn}
\affiliation{%
    \institution{Beijing University of Posts and Telecommunications
    }
    \streetaddress{No.10 Xitucheng Road}
    \state{Beijing}
    \country{China}
}

\author{Xumeng Gong}
\email{xumeng1141@bupt.edu.cn}
\affiliation{%
    \institution{Beijing University of Posts and Telecommunications}
    \streetaddress{No.10 Xitucheng Road}
    \state{Beijing}
    \country{China}
}

\author{Chuan Shi}
\authornote{Corresponding author}
\email{shichuan@bupt.edu.cn}
\affiliation{%
    \institution{Beijing University of Posts and Telecommunications
    }
    \streetaddress{No.10 Xitucheng Road}
    \state{Beijing}
    \country{China}
}
\author{Philip S. Yu}
\email{psyu@cs.uic.edu}
\affiliation{%
    \institution{UNIVERSITY OF ILLINOIS AT CHICAGO
    }
    \state{Chicago}
    \country{United States}
}
\renewcommand{\shortauthors}{ Cheng Yang, Xumeng Gong, Chuan Shi and Philip S. Yu}
\begin{abstract}
Recent years have witnessed the success of heterogeneous graph neural networks (HGNNs) in modeling heterogeneous information networks (HINs). In this paper, we focus on the benchmark task of HGNNs, \ie node classification, and empirically find that typical HGNNs are not good at predicting the label of a test node whose receptive field (1) has few training nodes from the same category or (2) has multiple training nodes from different categories. A possible explanation is that their message passing mechanisms may involve noises from different categories, and cannot fully explore task-specific knowledge such as the label dependency between distant nodes. Therefore, instead of introducing a new HGNN model,  we propose a general post-training framework that can be applied on any pretrained HGNNs to further inject task-specific knowledge and enhance their prediction performance. Specifically, we first design an auxiliary system that estimates node labels based on (1) a global inference module of multi-channel label propagation and (2) a local inference module of network schema-aware prediction. The mechanism of our auxiliary system can complement the pretrained HGNNs by providing extra task-specific knowledge. During the post-training process, we will strengthen both system-level and module-level consistencies to encourage the cooperation between a pretrained HGNN and our auxiliary system. In this way, both systems can learn from each other for better performance. In experiments, we apply our framework to four typical HGNNs. Experimental results on three benchmark datasets show that compared with pretrained HGNNs, our post-training framework can enhance Micro-F1 by  a relative improvement of $3.9\%$ on average. Code, data and appendix are available at \url{https://github.com/GXM1141/HGPF}.
\end{abstract}



\maketitle
\section{Introduction}
\begin{figure}[htb]
\centering
\subfigure[Close (red) v.s. Far (yellow)]{
\label{global observation}
\includegraphics[width = 0.46\columnwidth]{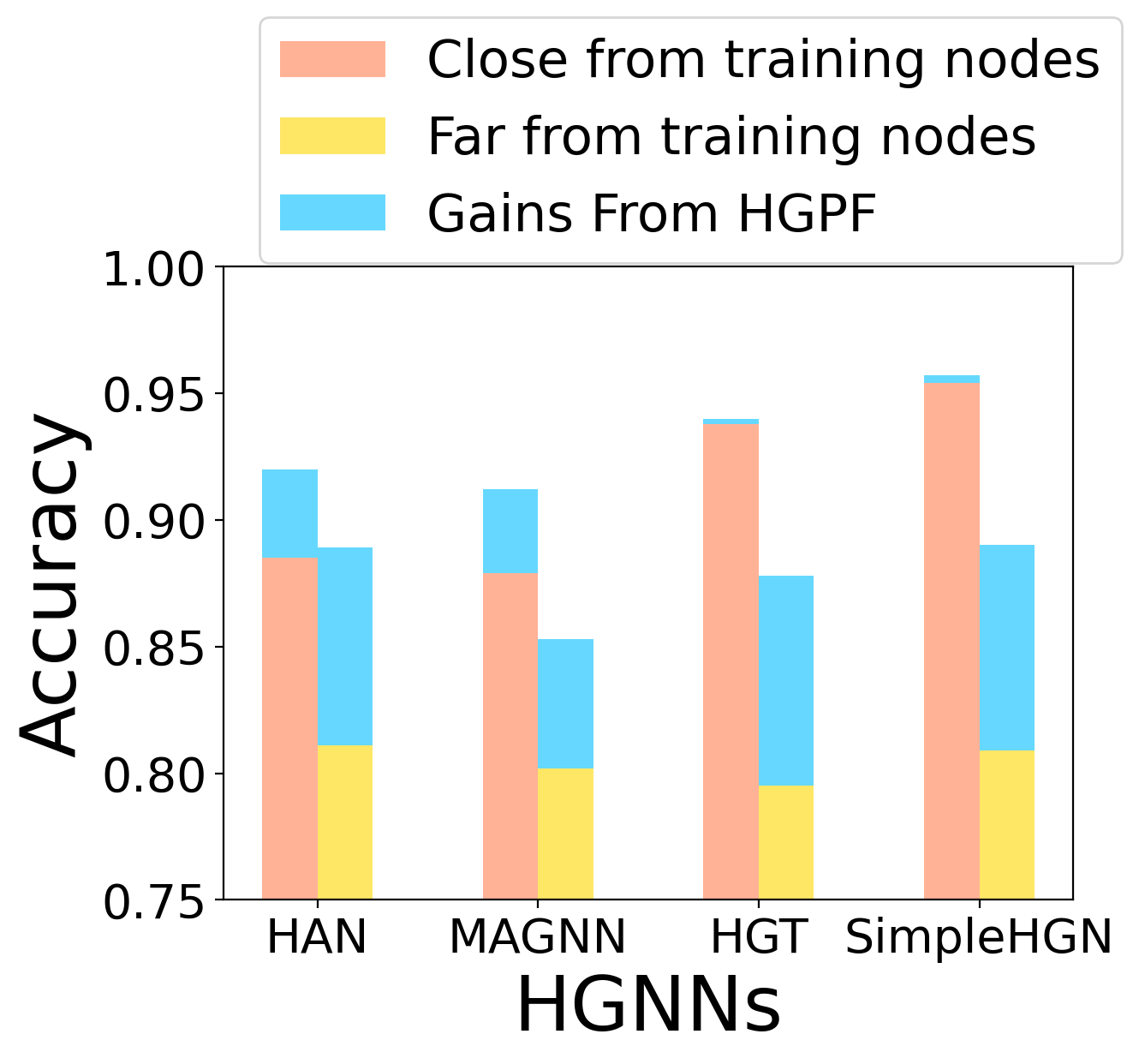}
}
\hspace{-0.1cm}
\subfigure[Same (red) v.s. Different (yellow)]{
\label{local observation}
\includegraphics[width = 0.48\columnwidth]{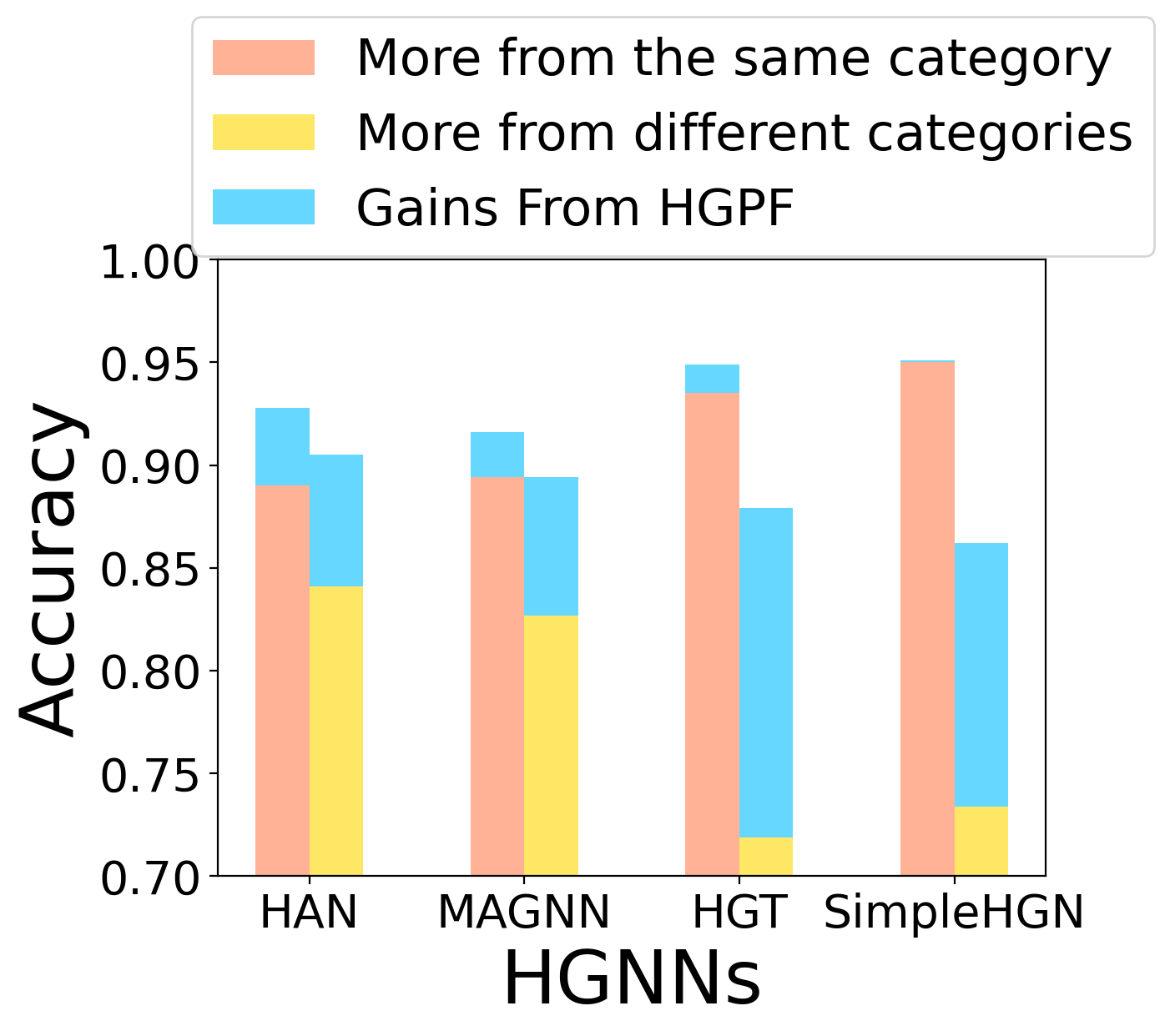}
}
\caption{Motivation verification on ACM dataset. The performance gaps between red and yellow columns indicate that all four HGNNs are not good at predicting the label of a test node: if (a) it's \textit{far} from training nodes of the same category or (b) it has more training nodes from \textit{different} categories than the same category in the receptive field. Blue columns are the performance gains from our post-training framework.}
\label{observation}
\end{figure}

A heterogeneous information network (HIN)~\cite{HIN,shi2021workshop,sun2022heterogeneous} can characterize the rich semantic relationship among a set of nodes and edges with different types. To effectively capture the semantics in an HIN, heterogeneous graph neural networks (HGNNs) were proposed in recent years~\cite{HAN,HetGNN,MAGNN,HGT}, and typical HGNNs will use meta-paths to enlarge the receptive field of each node in message passing~\cite{gilmer2017neural}. The success of HGNNs facilitated the development of HIN-based applications such as recommendation system~\cite{fan2019metapath,xu2019relation,lu2020meta, zheng2021multi} and malware detection system~\cite{fan2018gotcha,hei2021hawk,hou2017hindroid}.

In this work, we focus on the benchmark task for evaluating HGNNs, \ie semi-supervised node classification on HINs. We empirically find that typical HGNNs are not good at predicting the label of a test node whose receptive field (1) has few training nodes from the same category or (2) has multiple training nodes from different categories. For example, we evaluate four typical HGNNs~\cite{HAN,MAGNN,HGT,SimpleHGN} based on the benchmark setting~~\cite{NSPHINE} of ACM dataset. For the first observation, we compute the average distance from training nodes of the same category for every test node, and divide the test nodes into two groups depending on whether the average distance is larger than the range of receptive field or not. For the second observation, we build two node groups depending on whether a node's receptive field has more training nodes from the same category than different categories. As shown in Figure~\ref{observation}, the prediction accuracies of all four HGNNs drop significantly for the nodes that (1) are farther from training nodes of the same category or (2) have more training nodes from different categories than the same category in the receptive field. Note that such nodes occupy around 50 percent of all nodes. A possible reason is that the message passing mechanisms of typical HGNNs, which highly depend on the range of receptive fields, may involve noises from different categories, and cannot fully explore task-specific knowledge such as the label dependency between distant nodes. Moreover, this drawback cannot be addressed by simply enlarging the receptive field of HGNNs, \ie stacking more message passing layers, since more noises may be involved and the over-smoothing issue will be intensified.


As the above limitations generally exist in typical HGNNs, instead of introducing a new HGNN model, we propose a general post-training framework that can be applied on any pretrained HGNNs. Our Heterogeneous Graph Post-training Framework (HGPF) aims to alleviate the aforementioned limitations and enhance the prediction performance. Specifically, we first design an auxiliary system which predicts node labels based on a complementary mechanism: (1) a global inference module will diffuse node labels to distant nodes by multi-channel label propagation (usually $5$ times farther than the receptive fields of typical HGNNs in our experiments); (2) a local inference module will predict node labels merely based on every node's network schema instance, and thus exclude the influence of meta-path-based neighbors from potentially different categories. During the post-training process, we will optimize system-level prediction consistency between pretrained HGNN and auxiliary system, and module-level prediction consistency between global and local modules. In this way, both systems can learn from each other, and our auxiliary system can complement HGNNs by injecting extra task-specific knowledge for better performance. After the post-training, either updated HGNN or learned auxiliary system can be used for prediction.

To fully evaluate our proposed framework HGPF, we conduct experiments on three benchmark datasets and test with four typical HGNN models, including HAN~\cite{HAN}, HGT~\cite{HGT}, Simple-HGN~\cite{SimpleHGN} and MAGNN~\cite{MAGNN}. Experimental results show that compared with a pretrained HGNN, both updated HGNN and learned auxiliary system can have consistent improvements by learning from each other, and the learned auxiliary system performs best among the three. In terms of Micro-F1, the relative improvement of learned auxiliary system against pretrained HGNN is $3.9\%$ on average. We also compare our system against a broad range of state-of-the-art (SOTA) graph algorithms, showing that HGPF is the current SOTA method on this task. 


Our contributions are summarized as follows:

$\bullet$ We focus on the semi-supervised node classification task, and point out a key limitation of typical HGNNs that they are not good at predicting the label of a node whose receptive field has few training nodes from the same category or has multiple training nodes from different categories. 
    
$\bullet$  We propose a general and novel post-training framework HGPF that can be applied on any pretrained HGNNs to alleviate the above limitation and improve prediction accuracies. Specifically, we design an effective auxiliary system complementary with typical HGNNs, and encourage the two systems to learn from each other for better performance. 
    
$\bullet$  Experimental results show that systems learned by HGPF consistently outperform pretrained HGNNs by a large margin, and achieve SOTA performance on all three benchmark datasets compared with a variety of baselines.




\section{Related works}
\textbf{Heterogeneous graph neural networks:} Recently, many researchers focus on developing HGNNs for HIN modeling. Specifically, HAN~\cite{HAN} employed two types of attention mechanism to learn the node-level and semantic-level structures. MAGNN~\cite{MAGNN} further considered the intermediate nodes along the meta-paths on the basis of HAN. RSHN~\cite{RSHN} constructed a coarsened line graph to take edge features into account, and employed message passing neural network~\cite{MPNN} to propagate information of both nodes and edges. HetGNN~\cite{HetGNN} used random walks with restart mechanism to find valuable neighbors for nodes. HetSANN~\cite{HetSANN} employed a type-specific graph attention mechanism to aggregate the direct neighbors of nodes without meta-paths. GTN~\cite{GTN} was proposed to learn useful meta-paths automatically for message passing. HGT~\cite{HGT} was a transformer-based HGNN for modeling web-scale HINs through a graph sampling method. In addition to pairwise proximity, NSHE~\cite{NSPHINE} considered high-order proximity based on network schema instances. HGSL~\cite{HGSL} was proposed to jointly perform HIN structure learning and GNN parameter learning for classification. HGNN-AC~\cite{HGNN-AC} employed attention mechanism to complete missing attributes with MAGNN~\cite{MAGNN} as its backbone. HGK-GNN~\cite{HGK-GNN} proposed Heterogeneous Graph Kernel (HGK) based on Mahalanobis distance, and incorporated it into HGNNs. Simple-HGN~\cite{SimpleHGN} designed a simple HGNN model based on GAT~\cite{GAT}, and is the SOTA HGNN model.

\textbf{Connections with relevant models:} There are a few works that can be applied on homogeneous GNNs to boost their performance, though these methods did not use the term ``post-training''. For example, GMNN~\cite{qu2019gmnn} adopted an EM framework to train two GCNs iteratively, which can be seen as a multi-stage post-training method. RDD~\cite{zhang2020reliable} trained and ensembled multiple GCNs with a distillation framework to get better performance.  CPF~\cite{yang2021extract} used a knowledge distillation framework to extract the knowledge of GNNs and utilize more prior knowledge. These approaches are only applicable to homogeneous graphs, and their motivations are also different from ours. In our experiments, we will adopt these methods as our baselines.

Note that our auxiliary system includes a global and a local inference module. In fact, there are also some HGNN methods emphasizing the usage of global and local information. But these methods are mainly proposed for unsupervised representation learning and their definitions of global/local information are quite different from ours. For example, HDGI~\cite{HDGI} minimized the local-global mutual information between node-level representation and meta-path based graph-level representation. HeCo~\cite{HeCo} conducted contrastive learning between meta-path and schema views of an HIN to learn node representations in an unsupervised manner. Besides, the key designs of our auxiliary system, such as the multi-channel label propagation, have never been explored in previous HGNNs. 

\section{Preliminaries}

\begin{figure}[htb]
\begin{minipage}[h]{0.48\linewidth}
\subfigure[An example of HIN]{
\label{HIN}
\includegraphics[scale=0.16]{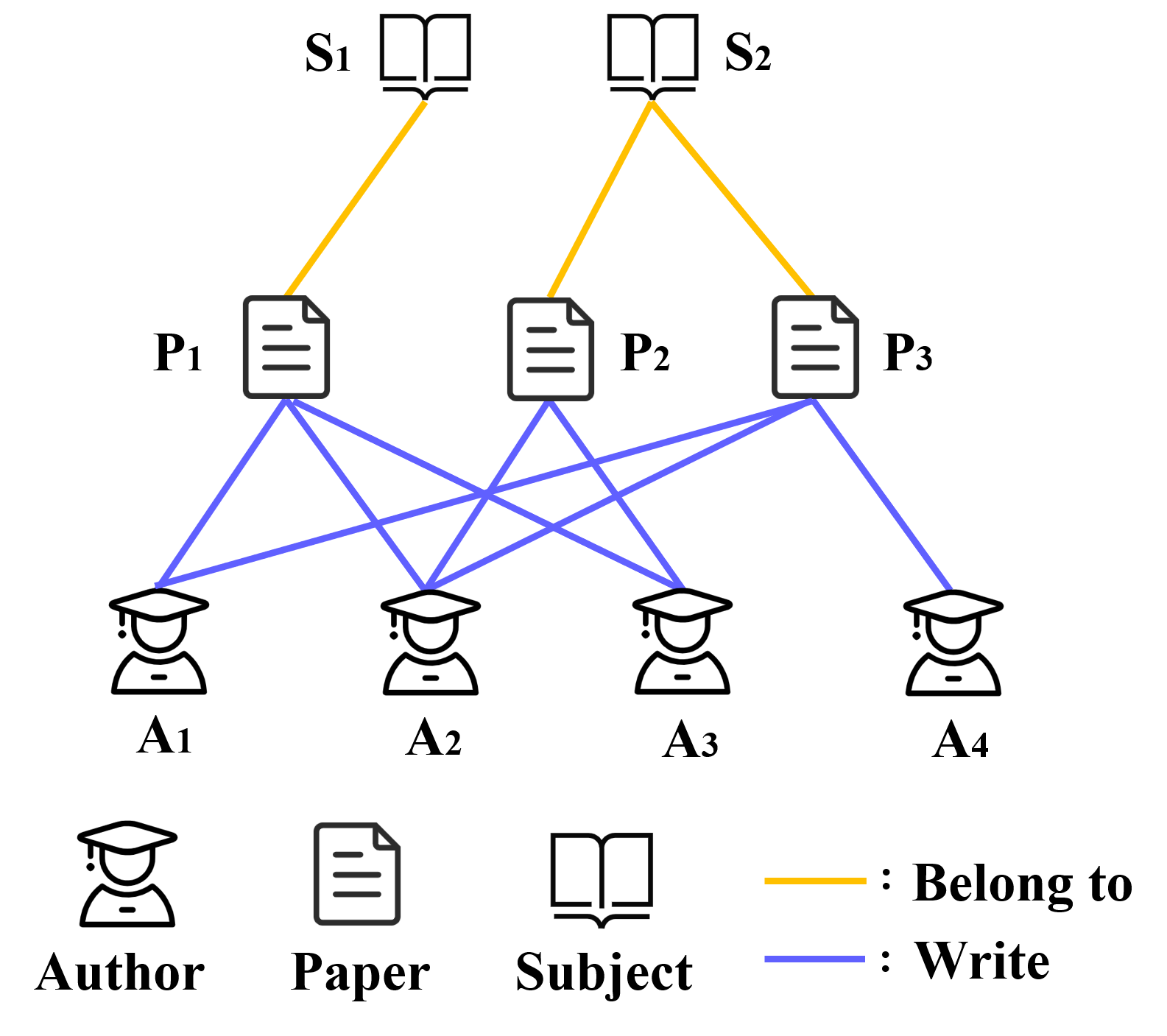}
}
\end{minipage}
\hfill
\hspace{-1cm}
\begin{minipage}[h]{0.40\linewidth}
\subfigure[Meta-paths]{
\includegraphics[scale=0.14]{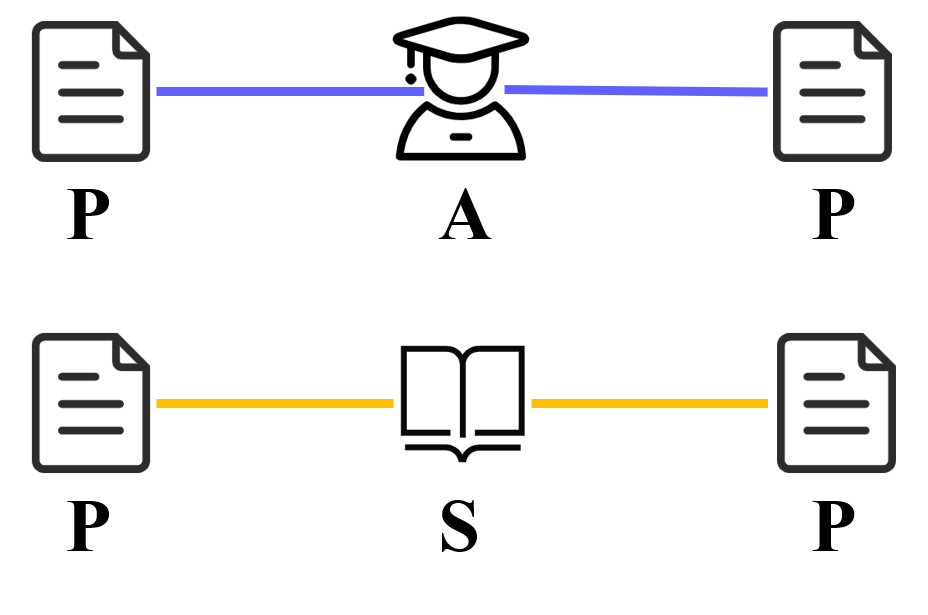}
\label{Meta-path}
}
\subfigure[Network schema]{
\includegraphics[scale=0.14]{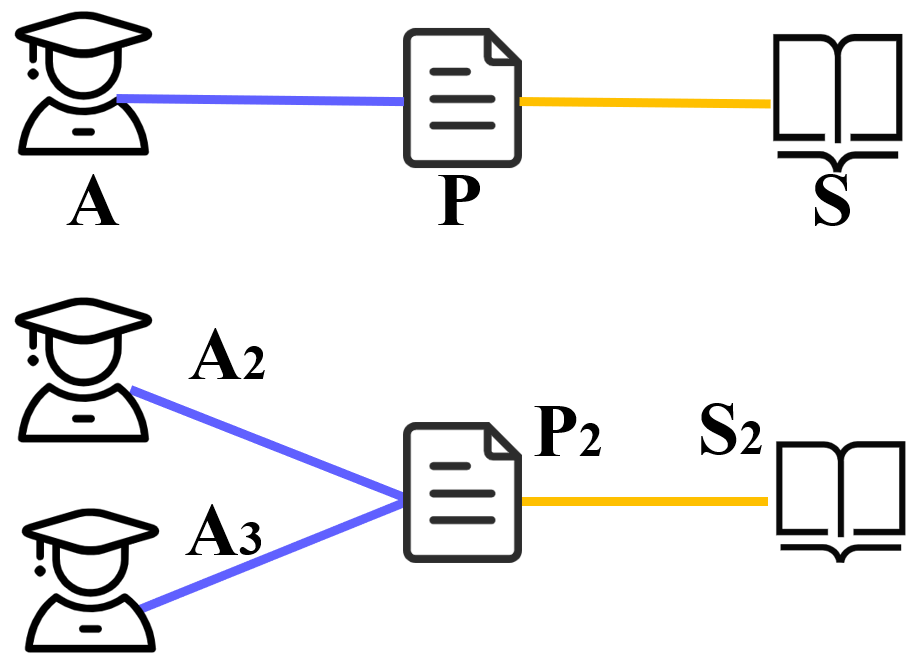}
\label{NetworkSchema}
}
\end{minipage}
\caption{A toy example of an HIN on ACM dataset~\cite{NSPHINE}.}
\label{pre_figure}
\end{figure}

\noindent\textbf{Definition 1: Heterogeneous Information Network.} A heterogeneous information network, denoted as $\mathcal{G} = \{\mathcal{V},\mathcal{E},\mathcal{T},\mathcal{R},\phi,\varphi\}$, is a special form of information networks, where $\mathcal{V}$ and  $\mathcal{E}$ denote the sets of nodes and edges, respectively. An HIN is also associated with a node type mapping function $\phi:\mathcal{V}\rightarrow\mathcal{T}$ and an edge type mapping function $\varphi:\mathcal{E}\rightarrow\mathcal{R}$, where $\mathcal{T}$ and $\mathcal{R}$ respectively denote the sets of node and edge types, with $|\mathcal{T}| + |\mathcal{R}| > 2$.

Fig.~\ref{HIN} illustrates an example of ACM dataset~\cite{NSPHINE}. The HIN has three types of nodes, including author (A), paper (P) and subject (S). There are also two types of edges (\ie ``write'' relation between author and paper, ``belong to'' relation between paper and subject). 

\noindent\textbf{Definition 2: Meta-path.} A meta-path $P$ is defined as a path in the form of ${T_1}{\stackrel{R_1}\longrightarrow}{T_2}{\stackrel{R_2}\longrightarrow}{\cdots}{\stackrel{R_l}\longrightarrow}{T_{l+1}}$ (abbreviated as $T_1T_2{\cdots}T_{l+1}$), which describes a composite relation $R = R_1{\hspace{0.2em} \circ \hspace{0.2em}}R_2{\hspace{0.2em} \circ \hspace{0.2em}}\cdots {\hspace{0.2em} \circ \hspace{0.2em}}R_l$ between node types $T_1$ and $T_{l+1}$, where $\circ$ denotes the composition operator on relations. We denote the meta-path set as $\mathcal{P}$ which contains all the meta-paths in an HIN. 

As shown in Fig.~\ref{Meta-path}, two papers can be connected via two kinds of meta-paths, \ie paper-author-paper (PAP) and paper-subject-paper (PSP). Different meta-paths represent different semantics in an HIN. For example, PAP connects two papers written by the same author, while PSP connects two papers from the same subject. 

\noindent\textbf{Definition 3: Network Schema.} The network schema $S_{\mathcal{G}} = (\mathcal{T}, \mathcal{R})$ is the blueprint of an HIN $\mathcal{G}$. Specifically, network schema is a directed graph defined on the set of node types $\mathcal{T}$, with edges as relations from the set of edge types $\mathcal{R}$. The network schema of ACM is shown in the upper half of Fig.~\ref{NetworkSchema}. In addition, the lower half of Fig.~\ref{NetworkSchema} presents a network schema instance, which is defined as a local structure matching the paradigm of network schema.

\noindent\textbf{Definition 4: Semi-supervised node classification on an HIN. } Given an HIN $\mathcal{G} = \{\mathcal{V},\mathcal{E},\mathcal{T},\mathcal{R},\phi,\varphi\}$ with node features $\mathcal{X}$, we aim to predict the labels for the nodes of a specific type $T \in \mathcal{T}$. We denote the  the set of nodes with type $T$ as the target node set $\mathcal{V}_T$. Each target node $v\in \mathcal{V}_T$ corresponds to a class label $y_v$ from the label set $\mathcal{Y}$. For the semi-supervised setting, the labels of nodes in labeled set $\mathcal{V}_L \subset{\mathcal{V}_T}$ are known, and the task is to predict the labels for the unlabeled nodes in $\mathcal{V}_U={{V}_T}\setminus \mathcal{V}_L$. Semi-supervised node classification is the most popular task for evaluating HGNN models~\cite{HAN,MAGNN,HGT,SimpleHGN}. 

\section{Methodology}
In this section, we will first introduce our design of the auxiliary system. Then we will present our post-training algorithm which strengthens the two levels of consistencies. Finally, we will have a discussion about the proposed framework.

\subsection{Formalization of HGNNs}
As our framework is agnostic to the neural architecture of HGNNs, we simply treat them as black boxes. Formally, an HGNN is a layered network architecture, which takes an HIN $\mathcal{G}$ and the corresponding feature matrix $\mathcal{X}$ as input to calculate the representation of each node. The representations of the last layer will be normalized by a softmax operator to output the label distribution predictions. In this paper, we denote an HGNN model parameterized by $\Theta$ as $f_{\Theta}$, where $f_{\Theta}(v)\in\mathbb{R}^{|\mathcal{Y}|}$ is the label distribution of node $v$ predicted by the HGNN model.

Then the model parameters $\Theta$ can be optimized by minimizing the prediction error on labeled node set $V_L$:
\begin{equation}
    	\underset{\Theta}{\min}{\sum_{v \in {\mathcal{V}_L}}{\mathcal{L}(f_{\Theta}(v), y_v)}},
    	\label{Eq:pre-train obejective}
\end{equation}
where $\mathcal{L}(\cdot, \cdot)$ denotes the loss function (\ie the cross entropy loss) between true label $y_v$ and predicted label $f_{\Theta}(v)$.

\subsection{Design of Auxiliary System}
\begin{figure*}[htb]
\centering
\includegraphics[scale = 0.25]{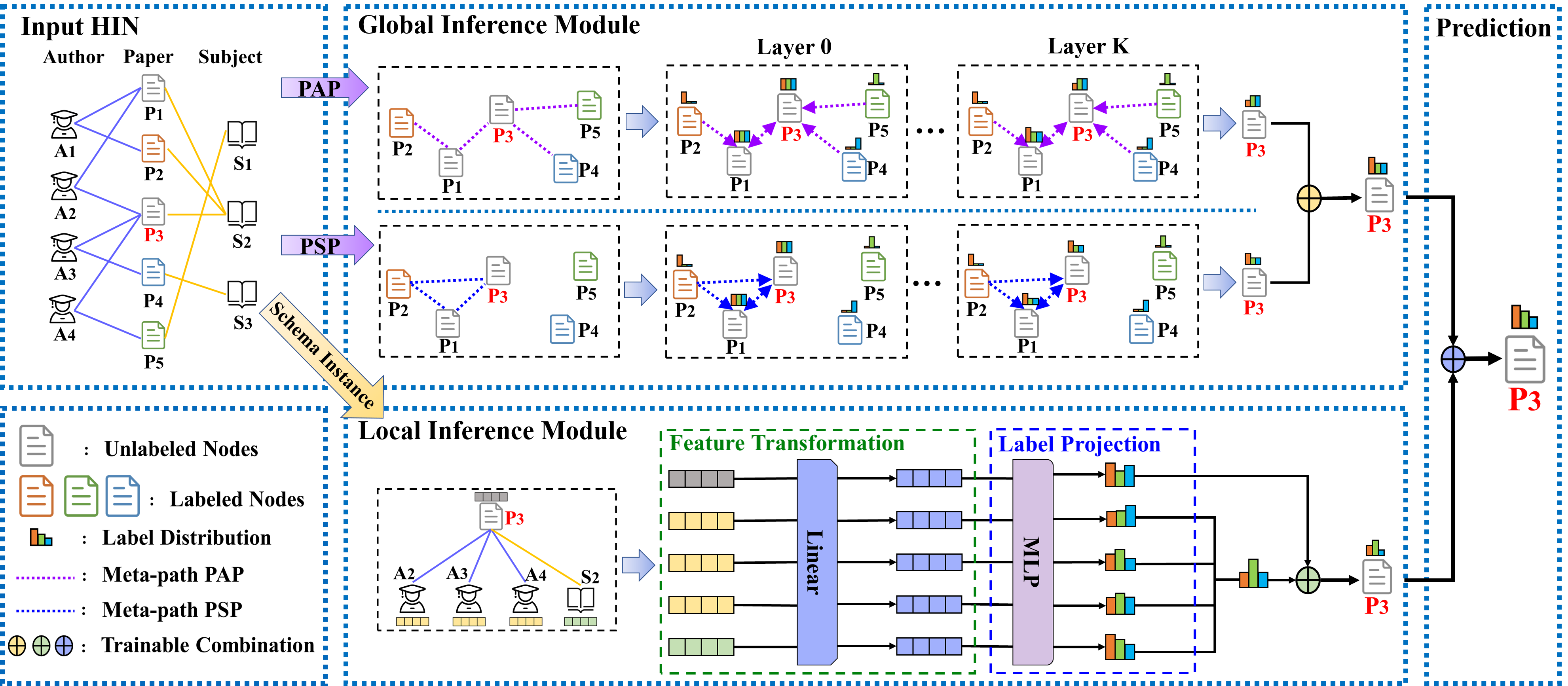}
\caption{The architecture of our auxiliary system. The global module based on multi-channel label propagation and the local module based on network schema-aware prediction will infer the label of node $P_3$ separately. Then their predictions will be recombined as the final prediction of the auxiliary system.}
\label{System2 Model}
\end{figure*}

In order to complement with typical HGNNs, our auxiliary system consists of two modules to characterize the global and local dependency. As shown in Fig.~\ref{System2 Model}, the two modules can infer the label distribution of a node separately, and their predictions can be further recombined by a learnable weighted average. Formally, we respectively denote the global and local inference modules as $g_{\Omega_1}^G$ and $g_{\Omega_2}^L$. The entire auxiliary system is represented as $g_{\Omega}$ where $\Omega=\{\Omega_1,\Omega_2\}$. Now we will introduce our global and local modules.


\subsubsection{Global Inference Module.} For global-level inference, we design a novel Multi-Channel Label Propagation (MCLP) process based on meta-paths. The basic assumption of MCLP is that nodes linked by a meta-path instance tend to have similar labels. In MCLP, labels will iteratively propagate from nodes to their meta-path based neighbors for inference. After several layers of propagation, the labels eventually propagate from labeled nodes to unlabeled ones as predictions. Note that an HIN usually has multiple meta-paths that define different semantic relations between nodes. Hence we will perform label propagation along different meta-paths in independent channels, and then recombine the predictions from different channels as the output of the global module. In our implementation, the global module will diffuse labels to the nodes around $5$ times farther than the receptive fields of typical HGNNs, and thus capture the label dependency between distant nodes.




Formally, we use $l_P^k$ to denote the predictions of MCLP based on meta-path $P\in\mathcal{P}$ after $k$ layers of propagation. We initialize the label prediction $l_P^0$ before propagation as follows:
\begin{equation}
    l^{0}_{P}(v)=\left\{
			\begin{aligned}
				(0,\dots 1,\dots 0)\in{\mathbb{R}^{|\mathcal{Y}|}},\ \ \ \ \ \ \ \ \ \ \ \forall{v\in\mathcal{V}_L}\\
				(\frac{1}{|\mathcal{Y}|},\dots\frac{1}{|\mathcal{Y}|},\dots\frac{1}{|\mathcal{Y}|})\in{\mathbb{R}^{|\mathcal{Y}|}},\ \forall{v\in\mathcal{V}_U} \\
			\end{aligned}
			\right.\\,
	\label{Eq:label initialize(LP)}
\end{equation}
where each labeled node corresponds to a one-hot label vector and all unlabeled nodes correspond to the uniform distribution.

We assume that the importance (\ie propagation intensities) of different neighbors to a node should be different. Thus, for each meta-path instance connecting node $u$ and $v$, we parameterize its propagation weight $w^P_{uv}\in[0,1]$ as follows:
\begin{equation}
    w^P_{uv}=\frac{\exp(s^P_{uv})}{\sum_{{u'}\in{\mathcal{N}^{P}_{v}}}\exp(s^P_{u'v})},
    \label{Eq:weight initialize(LP)}
\end{equation}
where $\mathcal{N}^P_v$ is the set of node $v$'s neighbors connected by meta-path instance of $P$, and $s^P_{uv}\in{\mathbb{R}}$ is a learnable parameter representing the propagation intensity between node $u$ and $v$ in channel $P$. 

At each layer of label propagation, we will update the label distribution predictions of unlabeled nodes, and fix those of labeled nodes as one-hot vectors. Specifically, the update function of node label predictions in the $k+1$-th layer can be formalized as:
\begin{equation}
    l^{k+1}_{P}(v)={\sum_{u\in{\mathcal{N}^{P}_{v}}}{w^{P}_{uv}l^{k}_{P}(u)}},
    \label{Eq:LP function(k->k+1)}
\end{equation}
where all the meta-path based neighbors of node $v$ will compete to propagate their labels to $v$.


Then we recombine the predictions from different channels to calculate the final predictions of MCLP as below:
\begin{equation}
    g_{\Omega_1}^G(v)={\sum_{{P} \in {\mathcal{P}}}}{{\alpha}^P_v} l_{P}^{K}(v),
    \label{Eq:global prediction}
\end{equation}
where $K$ is the number of layers in MCLP, ${\alpha}^P_v\in[0,1]$ is a trainable weight parameter for each pair of node $v$ and meta-path $P$ with ${\sum_{{P} \in {\mathcal{P}}}}{{\alpha}^P_v}=1$, and $g_{\Phi_1}^G(v)$ denotes the final predictions for node $v$ by the global inference module. 

\subsubsection{Local Inference Module.} We also design a local-level module to characterize each node's local neighborhood. The local module will predict node labels merely based on every node's network schema instance, and thus exclude the influence of meta-path-based neighbors from potentially different categories.


According to the network schema proximity~\cite{NSPHINE}, all the nodes with different types in a network schema instance tend to be similar. To take advantage of this proximity, we assume that besides the nodes with type $T$, all the nodes with other types can also be projected into the same label space $\mathcal{Y}$. Then the nodes in a network schema instance will have similar labels. Therefore, in order to help predict the label of node $v$, we will average the label distributions of all the nodes that are in the same network schema instances with $v$ for assistance.


Formally, we first apply a node type-specific transformation to project the features of different types of nodes into the same space:
\begin{equation}
    \mathbf{h}_{u}={W_{\phi(u)}\cdot\mathbf{x}_{u}}, \forall{u\in{\mathcal{V}}},
    \label{Eq:feature projection}
\end{equation}
where $\mathbf{h}_{u}$ is the projected feature of node $u$, $\mathbf{x}_{u}$ is the original feature of $u$, $\phi(u)$ is the type of node $u$, and $W_{\phi(u)}$ is the type-specific projection matrix.

Afterward, each node $u$ will be projected into a label distribution $p_u\in\mathbb{R}^{|\mathcal{Y}|}$ by
\begin{equation}
    p_u=\operatorname{softmax}(\operatorname{MLP}(\mathbf{h}_u)),
    \label{Eq:label_projection}
\end{equation}
where $\operatorname{MLP}(\cdot)$ is a multi-layer perceptron.

Then we denote the set of nodes that are in the same network schema instances with $v$ as ${\mathcal{N}_v}$. The labels of every node $u\in \mathcal{N}_v$ will be averaged and then combined with $p_v$ as the label distribution of node $v$ predicted by the local module:
\begin{equation}
    g_{\Omega_2}^L(v)=\beta_{v}p_v+(1-\beta_{v})\frac{\sum_{u\in{{\mathcal{N}_v}}} p_u}{|{\mathcal{N}_v|}},
    \label{Eq:Local prediction}
\end{equation}
where $\beta_{v}\in [0, 1]$ is a node-specific trainable parameter to balance the two parts of predictions. 



\subsubsection{The Combination of Global and Local Modules.}
Finally, the label distribution of node $v$ predicted by the auxiliary system is computed as
\begin{equation}
    g_{\Omega}(v)=\gamma_{v}g_{\Omega_1}^G(v) + (1-\gamma_{v})g_{\Omega_2}^L(v),
    \label{Eq:Combination of Global and Local}
\end{equation}
where $\gamma_{v}\in [0, 1]$ is a node-specific trainable parameter to balance the global module and local module. The pseudo code of our auxiliary system is provided in Appendix~\ref{code}.

\subsection{Post-training Algorithm}
To facilitate the cooperation of the two systems, we propose to optimize system-level and module-level consistencies, \ie minimizing the prediction gap between HGNN $f_{\Theta}$ and auxiliary system $g_\Omega$, as well as the gap between global module $g_{\Omega_1}^G$  and local module $g_{\Omega_2}^L$.

Given a pretrained HGNN model learned by its original training objective in Eq.~\ref{Eq:pre-train obejective}, we will then optimize the two systems alternatively. Formally, we will update parameter $\Omega=\{\Omega_1,\Omega_2\}$ in the auxiliary system by
\begin{equation}
    	\underset{\Omega}{\min}{\sum_{v \in {\mathcal{V}_U}}{\operatorname{dist}(f_{\Theta}(v), g_\Omega(v))}}+\lambda{{\operatorname{dist}(g_{\Omega_1}^G(v), g_{\Omega_2}^L(v))}},
    	\label{Eq:Cognitive System objective}
\end{equation}
where $\operatorname{dist}(\cdot, \cdot)$ denotes the distance function between two label distributions such as KL divergence or Euclidean distance, and $\lambda$ is a hyper-parameter. We empirically set $\lambda=0.3$ for all our experiments.

Afterward, we will update parameter $\Theta$ of HGNN by
\begin{equation}
    \underset{\Theta}{\min}{\sum_{v \in {\mathcal{V}_U}}{\operatorname{dist}(f_{\Theta}(v), g_\Omega(v))}}+{\sum_{v \in {\mathcal{V}_L}}{\mathcal{L}(f_{\Theta}(v), y_v)}},
    \label{Eq:Perceptive System objective}
\end{equation}
where the second term is the prediction error on labeled set to avoid trivial solutions, \eg both systems always predict a specific label $y\in \mathcal{Y}$.

By iteratively optimizing the two systems, they can learn from each other and both achieve better generalization ability. Note that both systems can be used for evaluation after training. We empirically find that the learned auxiliary system has better prediction accuracies. We name our overall framework as Heterogeneous Graph Post-training Framework, abbreviated as \modelname{}. The pseudo code of \modelname{} is provided in Appendix~\ref{code}. Note that the unlabeled node set ${\mathcal{V}}_{U}$ is further divided into the validation set ${\mathcal{V}}_{D}$ for model selection and test set ${\mathcal{V}}_{S}$ for final evaluation. We empirically set the maximum number of iterations and epochs $N=5$ and $M=150$ for all experiments. 



\subsection{Discussion}
\label{sec:discuss}
\subsubsection{Computational Complexity.} The time and space complexity of our framework is linear to the scale of an HIN, \ie the number of nodes, edges, features and meta-path instances. In fact, the training of \modelname{} is very time efficient. For example, if we apply \modelname{} on HAN~\cite{HAN}, the running time of \modelname{} on ACM dataset is about $5$ minutes with a single GPU device of GeForce RTX 3090. 

\subsubsection{System Complementarity.} Existing HGNN models can only have one or two stacked message passing layers to avoid the over-smoothing issue. In contrast, the MCLP process of our auxiliary system can be stacked for $8\sim 10$ layers to achieve better results. Thus, our auxiliary system can utilize more global information than typical HGNNs. Besides, many meta-path based HGNN models such as HAN~\cite{HAN} failed to consider the relationship between nodes with different types but in the same network schema instances. Hence the local module of our auxiliary system can also provide complementary knowledge with HGNNs.

\subsubsection{Model Compatibility.} Our proposed framework is agnostic to the architecture of HGNNs, and thus can be integrated with any HGNN models for implementation. After the post-training phase, we can offer more accurate predictions and have a high compatibility in improving HGNN models.




%

\section{Experiments}
In this section, we will conduct experiments to answer the following research questions: $\bullet$ \textbf{RQ1}: Can \modelname{} improve HGNNs on semi-supervised node classification task? $\bullet$ \textbf{RQ2}: How does \modelname{} perform compared with SOTA graph algorithms? $\bullet$ \textbf{RQ3}: How does \modelname{} perform under different settings, \ie training ratios, ablated models, and hyper-parameters? $\bullet$ \textbf{RQ4}: How about the interpretability of learned parameters in the auxiliary system? 

\subsection{Datasets}
We adopt three benchmark datasets including ACM~\cite{NSPHINE}, DBLP~\cite{MAGNN} and IMDB~\cite{MAGNN} for evaluation.
The detailed statistics and descriptions are listed in Appendix~\ref{datasets}. For semi-supervised node classification task, we randomly choose $20$ or $50$ labeled nodes per class as the training set, $50$ nodes per class as the validation set, and the remaining nodes as the test set. Following previous HGNNs~\cite{HAN,SimpleHGN,MAGNN}, we also employ Micro-F1 and Macro-F1 as evaluation metrics. For auxiliary experiments in Section~\ref{sec:aux}, we only report the Micro-F1 metric on ACM for brevity. For the analysis related to interpretability in Section~\ref{weightob}, we will focus on DBLP, which has richer semantics with the largest numbers of node types, edge types and meta-paths.
\subsection{Experimental Settings}
Here we only present some key settings and more detailed experimental settings are provided in Appendix~\ref{other setting}.
\subsubsection{System Settings} 
To prove the effectiveness of \modelname{}, we explore four typical HGNN models, including HAN \cite{HAN}, HGT \cite{HGT}, Simple-HGN \cite{SimpleHGN} and MAGNN \cite{MAGNN}. Since the implementation with MAGNN is much slower than that with the other three HGNNs, we only evaluate MAGNN in main experiments. All the HGNNs are carefully tuned according to our dataset splits, and a sufficient training can be guaranteed during the pretraining phase. For the hyper-parameter setting of our auxiliary system, we explore MCLP layers $K$ from $6$ to $14$, and use a 2-layer MLP with hidden size as $128$ in our local module.


\subsubsection{Optimization Settings.} We will alternatively run the two systems for 150 epochs in each iteration, and choose the best epoch and iteration according to the performance on the validation set.

\subsubsection{Models for Comparisons.} In the experiments, we will consider three models in each group of comparison:

$\bullet$ \textbf{Pretrain}. The pretrained HGNN, \ie a standard HGNN. 

$\bullet$ \textbf{\modelname{}$_\text{self}$}. A variant of \modelname{} where the auxiliary system is replaced by the same HGNN without sharing parameters.

$\bullet$ \textbf{\modelname{}}. The learned auxiliary system \modelname{}$_\text{AS}$ in our \modelname{}. We omit the subscript of AS for brevity. Note that the performance of updated HGNN \modelname{}$_\text{HGNN}$ will be discussed in RQ3, and \modelname{} will refer to learned auxiliary system unless specified.


\subsection{Performance on Node Classification (RQ1)}
Experimental results on three datasets with four HGNNs are presented in Table \ref{Tab:classification results of HAN and HGT} and \ref{Tab:classification results of SimpleHGN and MAGNN}. We bold the best result in each group of comparison, and have the following observations:

(1) Comparing \modelname{} with Pretrain, we can find that our learned auxiliary system consistently outperforms the pretrained HGNN in all 48 groups of comparisons. The relative improvements of Micro-F1 and Macro-F1 are respectively $3.90\%$ and $3.94\%$ on average. Also, the standard derivation of \modelname{} is about $\pm 0.3$ and thus the improvements are significant. Hence our proposed framework can be successfully integrated with all four HGNNs, and offer more accurate predictions.

(2) Comparing \modelname{} with \modelname{}$_\text{self}$ where the auxiliary system has the same architecture with pretrained HGNN, we can see that our carefully designed auxiliary system has better performance in 47 out of 48 groups of comparisons. The relative improvements of Micro-F1 and Macro-F1 are respectively $1.87\%$ and $2.04\%$ on average. This observation shows that our auxiliary system can better complement with typical HGNNs, which demonstrates the effectiveness of our model design.

(3) Comparing \modelname{}$_\text{self}$ with Pretrain, though two systems have the same architecture, we can still get some improvement by our optimization framework. A possible reason is that the two systems are optimized into different local minimums, and thus can also complement with each other to some extent. Different from ensemble, the learned system won't increase the complexity at test stage. 

\begin{table*}
  \centering
  \caption{Classification performance with HAN~\cite{HAN} and HGT~\cite{HGT}.}
    \begin{tabular}{c|c|cc|cc|cc||cc|cc|cc}
    \toprule[1pt]
    \multicolumn{2}{c|}{\multirow{2}[4]{*}{\textbf{Models}}} & \multicolumn{6}{c||}{\textbf{HAN}}             & \multicolumn{6}{c}{\textbf{HGT}} \\
\cmidrule{3-14}    \multicolumn{2}{c|}{} & \multicolumn{2}{c|}{\textbf{Pretrain}} & \multicolumn{2}{c|}{\textbf{\modelname{}$_\text{self}$}} & \multicolumn{2}{c||}{\textbf{\modelname{}}} & \multicolumn{2}{c|}{\textbf{Pretrain}} & \multicolumn{2}{c|}{\textbf{\modelname{}$_\text{self}$}} & \multicolumn{2}{c}{\textbf{\modelname{}}} \\
    \midrule
    \multicolumn{2}{c|}{\textbf{\# Labeled Nodes}} & \textbf{20} & \textbf{50} & \textbf{20} & \textbf{50} & \textbf{20} & \textbf{50} & \textbf{20} & \textbf{50} & \textbf{20} & \textbf{50} & \textbf{20} & \textbf{50} \\
    \midrule
    \multirow{2}[2]{*}{\textbf{ACM}} & \textbf{Micro-F1} & 0.8826  & 0.8838  & 0.8949  & 0.9091  & \textbf{0.9163 } & \textbf{0.9271 } & 0.8693  & 0.8701  & 0.8835  & 0.8852  & \textbf{0.9173 } & \textbf{0.9177 } \\
          & \textbf{Macro-F1} & 0.8785  & 0.8864  & 0.8901  & 0.9054  & \textbf{0.9165 } & \textbf{0.9280 } & 0.8679  & 0.8699  & 0.8827  & 0.8807  & \textbf{0.9173 } & \textbf{0.9174 } \\
    \midrule
    \multirow{2}[2]{*}{\textbf{DBLP}} & \textbf{Micro-F1} & 0.9092  & 0.9217  & 0.9251  & 0.9300  & \textbf{0.9280 } & \textbf{0.9349 } & 0.8941  & 0.9256  & 0.9011  & 0.9317  & \textbf{0.9084 } & \textbf{0.9342 } \\
          & \textbf{Macro-F1} & 0.9038  & 0.9165  & 0.9220  & 0.9242  & \textbf{0.9258 } & \textbf{0.9287 } & 0.8871  & 0.9229  & 0.8925  & 0.9239  & \textbf{0.8995 } & \textbf{0.9275 } \\
    \midrule
    \multirow{2}[2]{*}{\textbf{IMDB}} & \textbf{Micro-F1} & 0.4581  & 0.4809  & 0.4629  & 0.5060  & \textbf{0.4879 } & \textbf{0.5149 } & 0.4600  & 0.5067  & 0.4672  & 0.5117  & \textbf{0.4724 } & \textbf{0.5286 } \\
          & \textbf{Macro-F1} & 0.4346  & 0.4817  & 0.4574  & 0.5059  & \textbf{0.4792 } & \textbf{0.5189 } & 0.4540  & 0.5123  & \textbf{0.4623}  & 0.5139  & 0.4611  & \textbf{0.5328 } \\
    \bottomrule[1pt]
    \end{tabular}%
  \label{Tab:classification results of HAN and HGT}%
\end{table*}%

\begin{table*}[htbp]
  \centering
  \caption{Classification performance with Simple-HGN~\cite{SimpleHGN} and MAGNN~\cite{MAGNN}.}
    \begin{tabular}{c|c|cc|cc|cc||cc|cc|cc}
    \toprule[1pt]
    \multicolumn{2}{c|}{\multirow{2}[4]{*}{\textbf{Models}}} & \multicolumn{6}{c||}{\textbf{Simple-HGN}}             & \multicolumn{6}{c}{\textbf{MAGNN}} \\
\cmidrule{3-14}    \multicolumn{2}{c|}{} & \multicolumn{2}{c|}{\textbf{Pretrain}} & \multicolumn{2}{c|}{\textbf{\modelname{}$_\text{self}$}} & \multicolumn{2}{c||}{\textbf{\modelname{}}} & \multicolumn{2}{c|}{\textbf{Pretrain}} & \multicolumn{2}{c|}{\textbf{\modelname{}$_\text{self}$}} & \multicolumn{2}{c}{\textbf{\modelname{}}} \\
    \midrule
    \multicolumn{2}{c|}{\textbf{\# Labeled Nodes}} & \textbf{20} & \textbf{50} & \textbf{20} & \textbf{50} & \textbf{20} & \textbf{50} & \textbf{20} & \textbf{50} & \textbf{20} & \textbf{50} & \textbf{20} & \textbf{50} \\
    \midrule
    \multirow{2}[2]{*}{\textbf{ACM}} & \textbf{Micro-F1} & 0.8816  & 0.8865  & 0.8945  & 0.8994  & \textbf{0.9179 } & \textbf{0.9216 } & 0.8776  & 0.8831  & 0.8917  & 0.9022  & \textbf{0.9157 } & \textbf{0.9179 } \\
          & \textbf{Macro-F1} & 0.8815  & 0.8881  & 0.8895  & 0.8943  & \textbf{0.9179 } & \textbf{0.9210 } & 0.8715  & 0.8824  & 0.8923  & 0.9014  & \textbf{0.9112 } & \textbf{0.9173 } \\
    \midrule
    \multirow{2}[2]{*}{\textbf{DBLP}} & \textbf{Micro-F1} & 0.9108  & 0.9253  & 0.9245  & 0.9315  & \textbf{0.9279 } & \textbf{0.9366 } & 0.9121  & 0.9223  & 0.9271  & 0.9322  & \textbf{0.9291 } & \textbf{0.9359 } \\
          & \textbf{Macro-F1} & 0.9026  & 0.9249  & 0.9152  & 0.9286  & \textbf{0.9177 } & \textbf{0.9316 } & 0.9056  & 0.9228  & 0.9234  & 0.9269  & \textbf{0.9252 } & \textbf{0.9295 } \\
    \midrule
    \multirow{2}[2]{*}{\textbf{IMDB}} & \textbf{Micro-F1} & 0.4698  & 0.5109  & 0.4798  & 0.5318  & \textbf{0.4925 } & \textbf{0.5412 } & 0.4518  & 0.5090  & 0.4877  & 0.5173  & \textbf{0.4962 } & \textbf{0.5292 } \\
          & \textbf{Macro-F1} & 0.4562  & 0.5141  & 0.4522  & 0.5296  & \textbf{0.4874 } & \textbf{0.5396 } & 0.4515  & 0.5119  & 0.4880  & 0.5204  & \textbf{0.4921 } & \textbf{0.5256 } \\
    \bottomrule[1pt]
    \end{tabular}%
  \label{Tab:classification results of SimpleHGN and MAGNN}%
\end{table*}%

\subsection{Comparison with SOTA GNNs (RQ2)}
\label{SOTA-exp}
\begin{figure*}[htb]
\centering
\subfigure[ACM]{
\label{ACM_SOTA}
\includegraphics[width = 0.5\columnwidth]{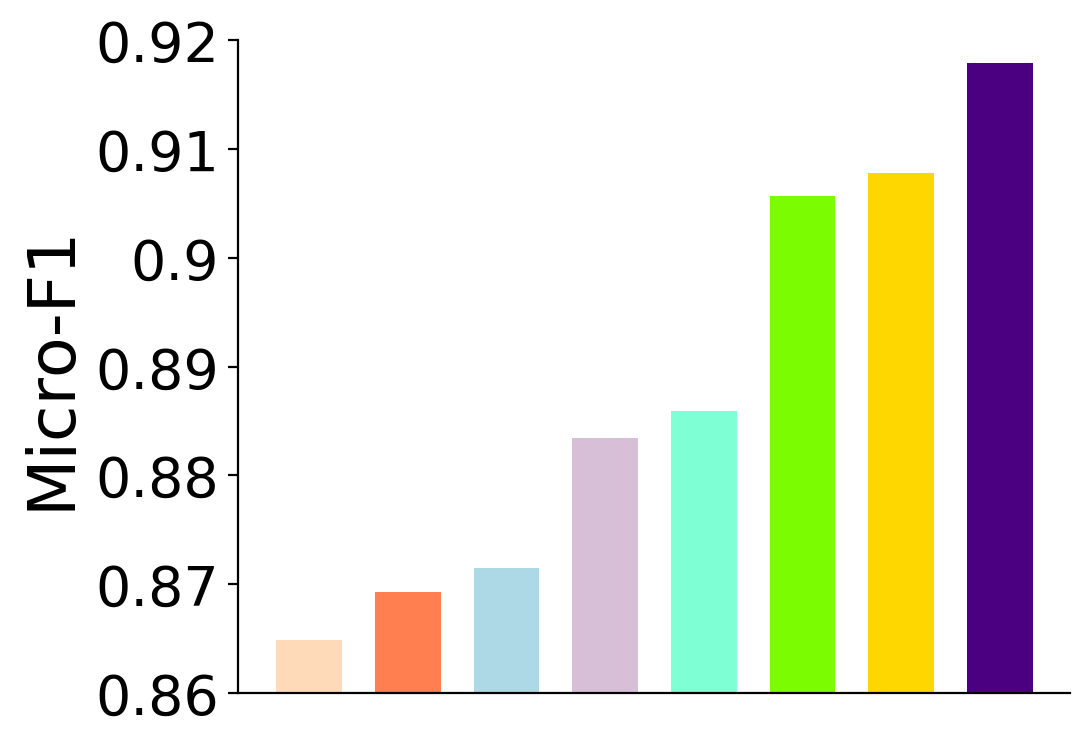}
}
\hspace{-0.1cm}
\subfigure[DBLP]{
\label{DBLP_SOTA}
\includegraphics[width = 0.51\columnwidth]{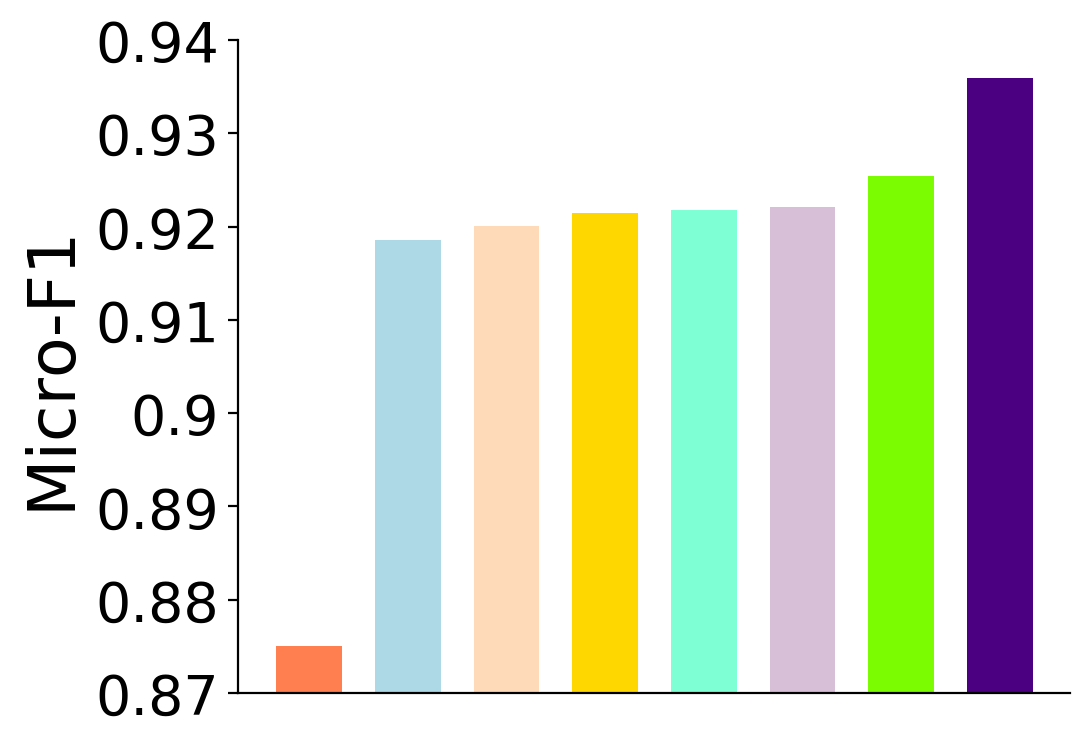}
}
\hspace{-0.1cm}
\subfigure[IMDB]{
\label{IMDB_SOTA}
\includegraphics[width = 0.51\columnwidth]{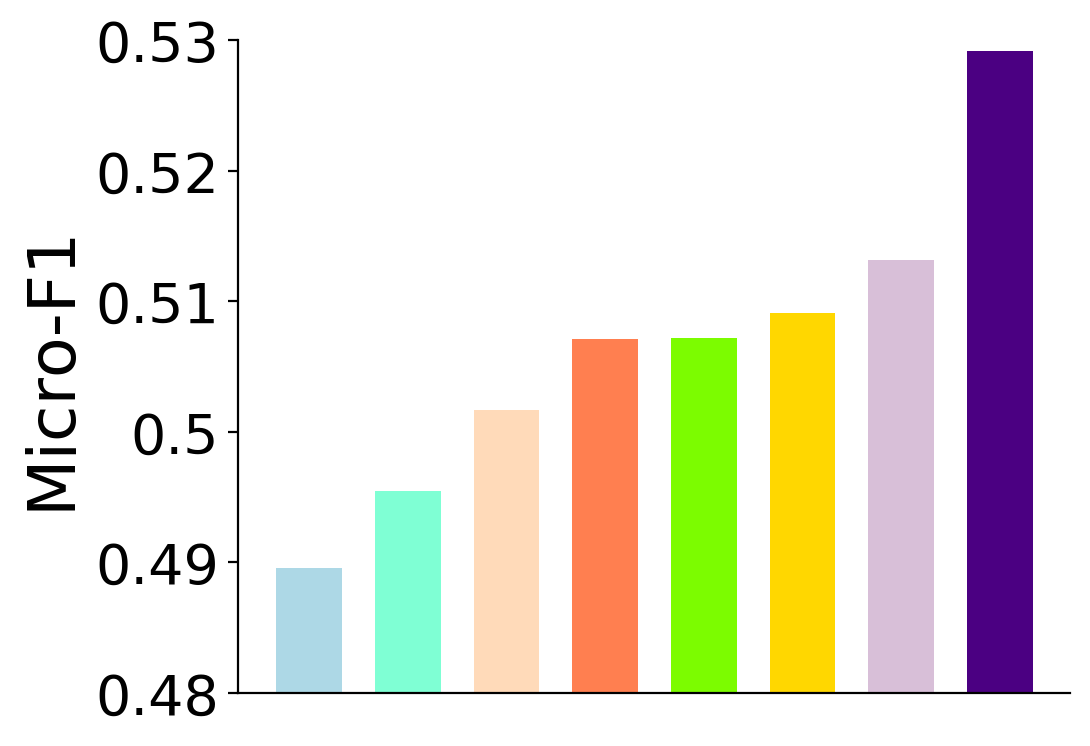}
}
\hspace{-0.05cm}
\subfigure{
\label{SOTA_legends}
\includegraphics[width = 0.25\columnwidth]{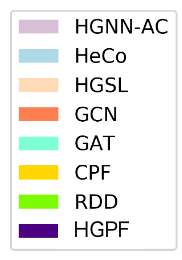}
}
\caption{Performance of \modelname{} and seven state-of-the-art GNNs and HGNNs on the three datasets.}
\label{SOTA}
\end{figure*}

Besides the four HGNNs used in our \modelname{}, in this subsection we conduct experiments to compare with more state-of-the-art GNNs, including two HGNNs (HGSL~\cite{HGSL}, HeCo~\cite{HeCo}) and three frameworks (RDD~\cite{zhang2020reliable}, CPF~\cite{yang2021extract}, HGNN-AC~\cite{HGNN-AC}) that can be applied on any GNNs. For a fair comparison, we use MAGNN~\cite{MAGNN} as the backbone in the three frameworks and our \modelname{}. Also, since a recent work~\cite{SimpleHGN} mentioned that typical homogeneous GNNs perform well for HINs in practice, we add GCN~\cite{GCN} and GAT~\cite{GAT} for comparison as well. Figure~\ref{SOTA} shows the comparison results on the three datasets with 50 labeled nodes per class. The relative improvements of Micro-F1 against best performed baselines on three datasets are respectively $1.10\%$, $1.13\%$, $3.12\%$, which demonstrates that our \modelname{} is the SOTA method for this task. Detailed settings of the seven baselines are provided in Appendix~\ref{other setting}.

\subsection{Performance under Different Settings (RQ3)}
\label{sec:aux}
\subsubsection{Analysis of Different Training Ratios}
\label{ACMs}
\begin{figure*}[htb]
\centering
\subfigure[HAN]{
\label{HAN_Ratio}
\includegraphics[width = 0.52\columnwidth]{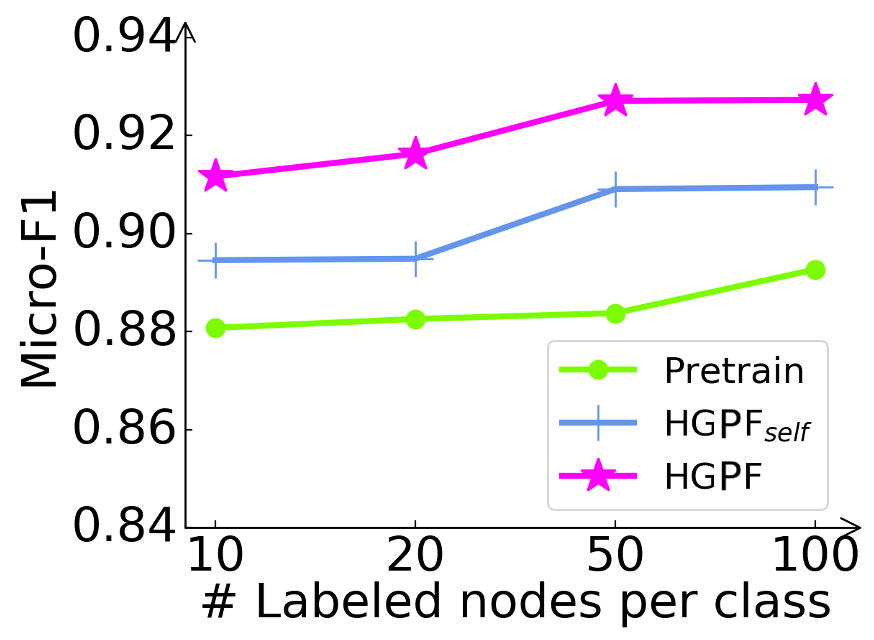}}
\hspace{+0.5cm}
\subfigure[HGT]{
\label{HGT_Ratio}
\includegraphics[width = 0.52\columnwidth]{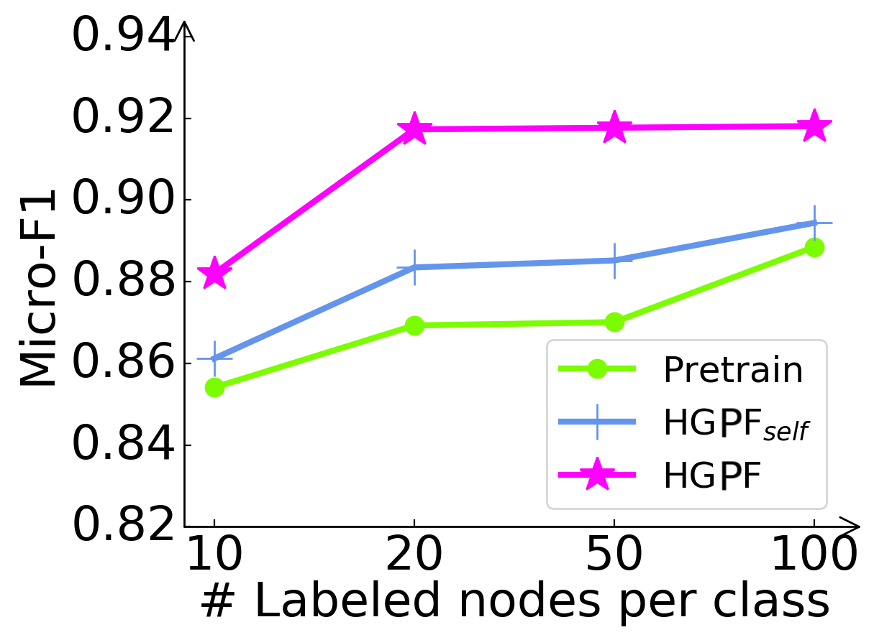}
}
\hspace{+0.5cm}
\subfigure[Simple-HGN]{
\label{SimpleHGN_Ratio}
\includegraphics[width = 0.52\columnwidth]{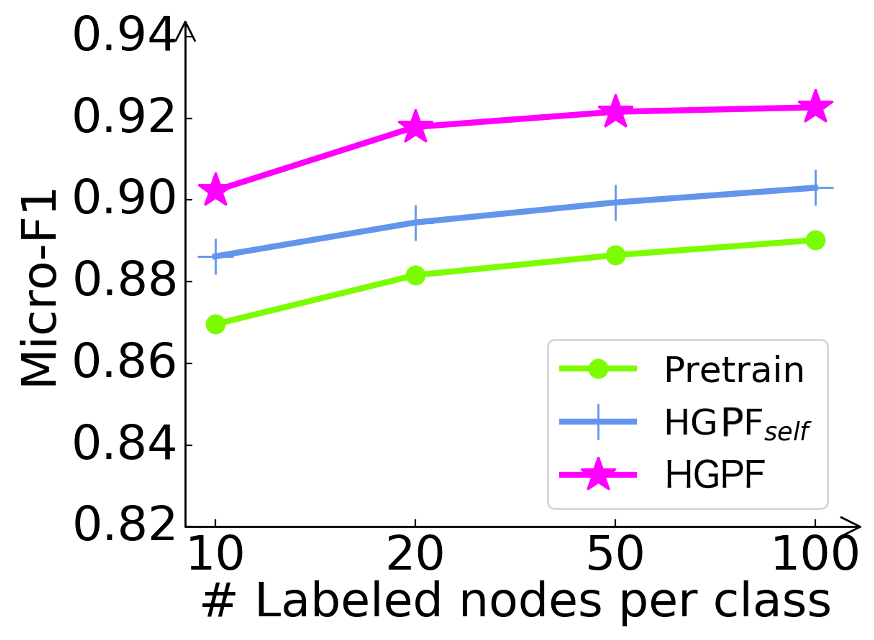}
}
\caption{Classification performance on ACM dataset under different training ratios.}
\label{Ratio}
\end{figure*}

In this subsection, we conduct experiments under different training ratios to further prove the robustness of our framework. Specifically, we report the classification results with 10, 20, 50, and 100 labeled nodes per class. From Figure \ref{Ratio}, we can see that the results of \modelname{} are consistently better than both the pretrained HGNN models and \modelname{}$_\text{self}$ when the number of labeled nodes increases. Note that \modelname{} can even outperform pretrained HGNNs with only $1/5$ or $1/10$ labeled data. For example, the performance of \modelname{} with 10 labeled nodes per class is better than that of pretrained HAN with 100 labeled nodes per class. Hence our framework is stable and effective with different training ratios. 

\subsubsection{Ablation Study of \modelname{}}
\begin{figure}[htb]
\centering
\subfigure[20 labeled nodes per class.]{
\label{20_Micro_Ablation}
\includegraphics[width = 0.45\columnwidth]{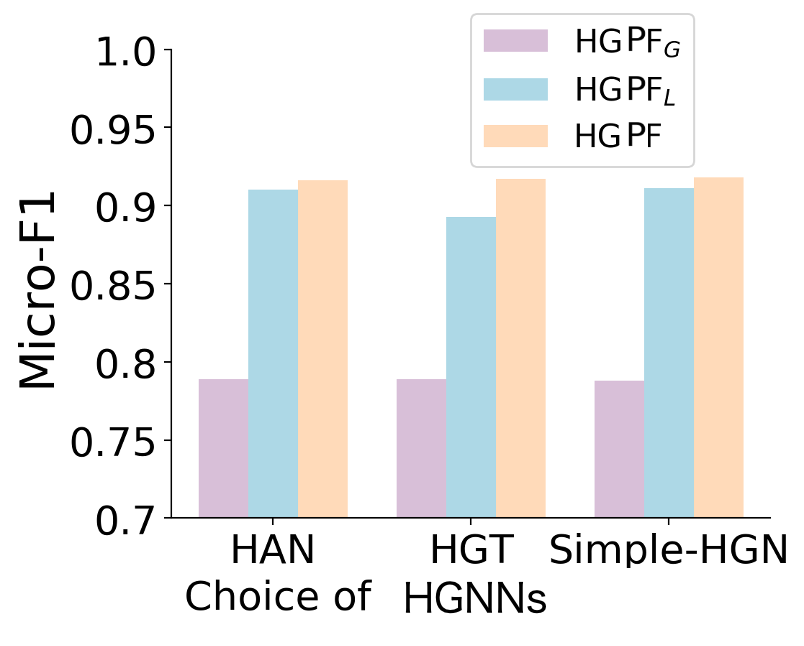}}
\hspace{-0.1cm}
\subfigure[50 labeled nodes per class.]{
\label{50_Micro_Ablation}
\includegraphics[width = 0.45\columnwidth]{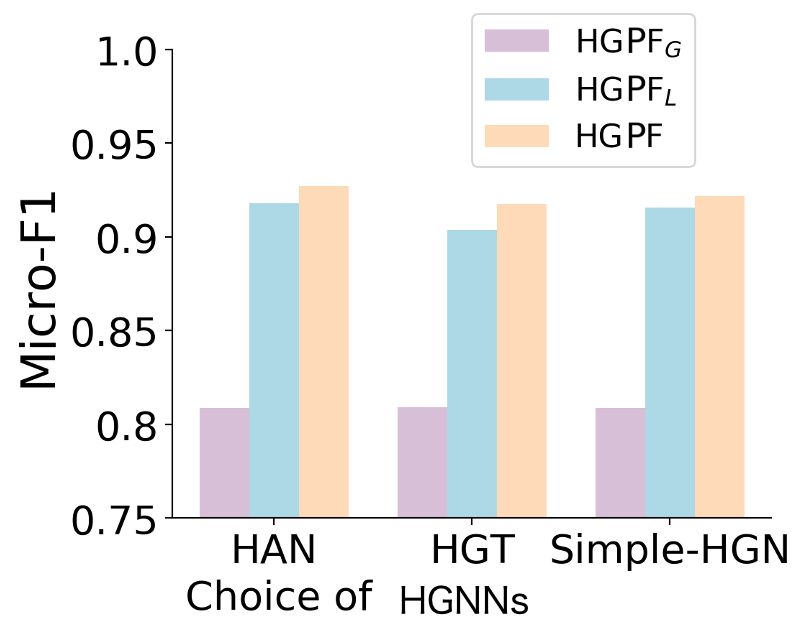}
}
\caption{Ablation study on global/local modules on ACM.}
\label{ablation}
\end{figure}

To further demonstrate the effectiveness of our global and local inference modules, we design two variants of \modelname{}:

 $\bullet$ \textbf{\modelname{}$_\text{G}$}: The variant of \modelname{} with only global inference module in the auxiliary system. 
 
 $\bullet$ \textbf{\modelname{}$_\text{L}$}: The variant of \modelname{} with only local inference module in the auxiliary system.

Experimental results on ACM dataset are shown in Figure \ref{ablation}. We can see that \modelname{} always has better performance than two ablated models. Compared with \modelname{}$_\text{L}$, \modelname{} has $1.23\%$ relative improvement on average. Hence, although \modelname{}$_\text{G}$ performs worst among the three, the global inference module is still an indispensable part of our auxiliary system. This experiment demonstrates the necessity of both global and local modules.

\subsubsection{Analysis of Hyper-parameters}
\label{ACMe}

\begin{figure}[htb]
\centering
\subfigure[20 labeled nodes per class]{
\label{20Iterations}
\includegraphics[width = 0.35\columnwidth]{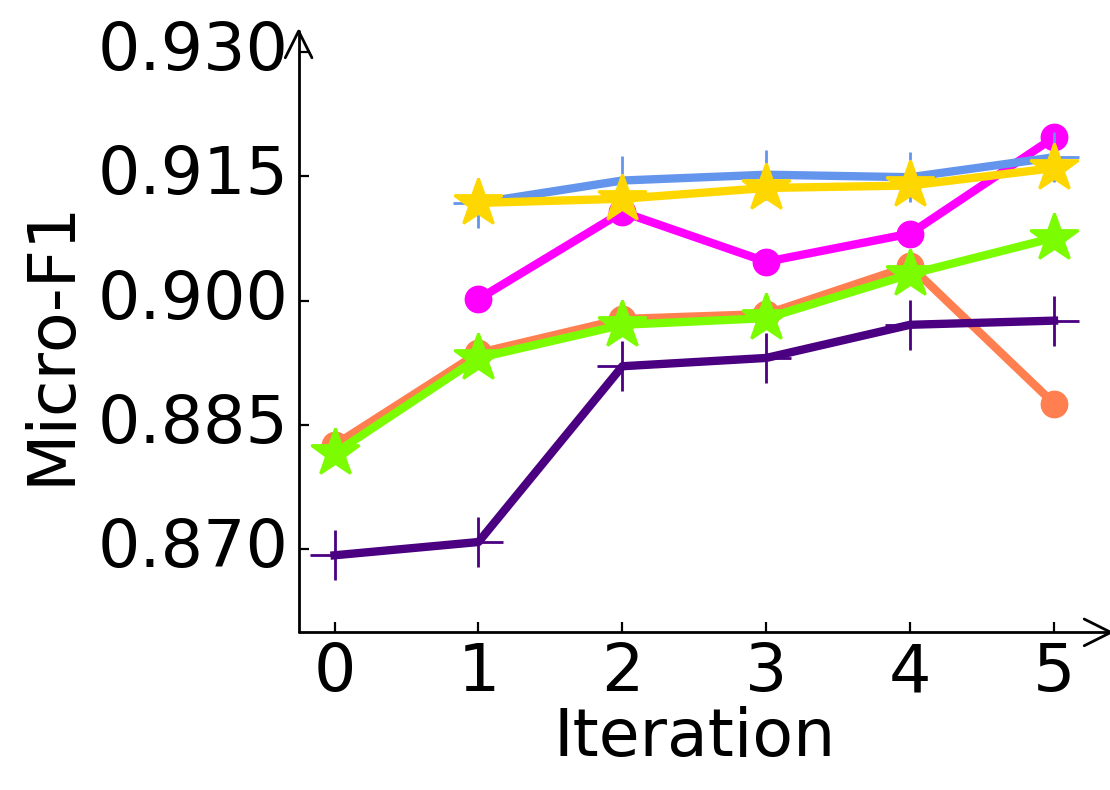}
}
\hspace{-0.1cm}
\subfigure[50 labeled nodes per class]{
\label{50Iterations}
\includegraphics[width = 0.35\columnwidth]{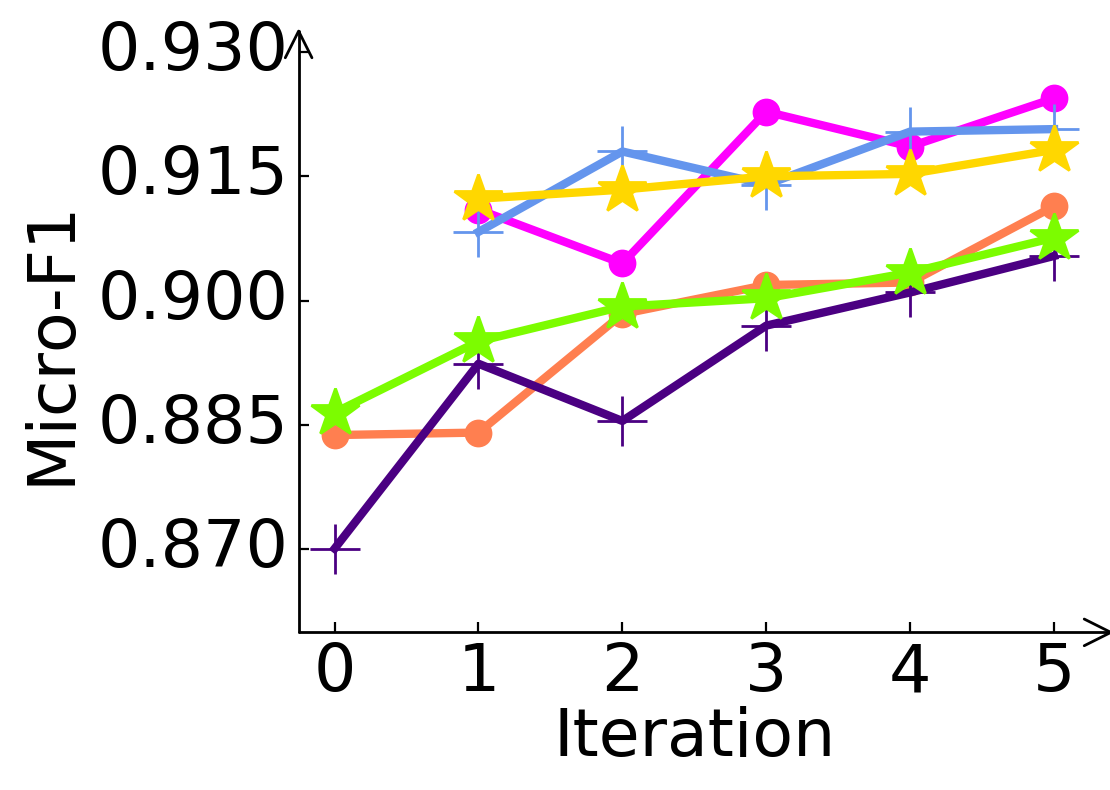}
}
\hspace{-0.1cm}
\subfigure{
\label{legend}
\includegraphics[width = 0.23\columnwidth]{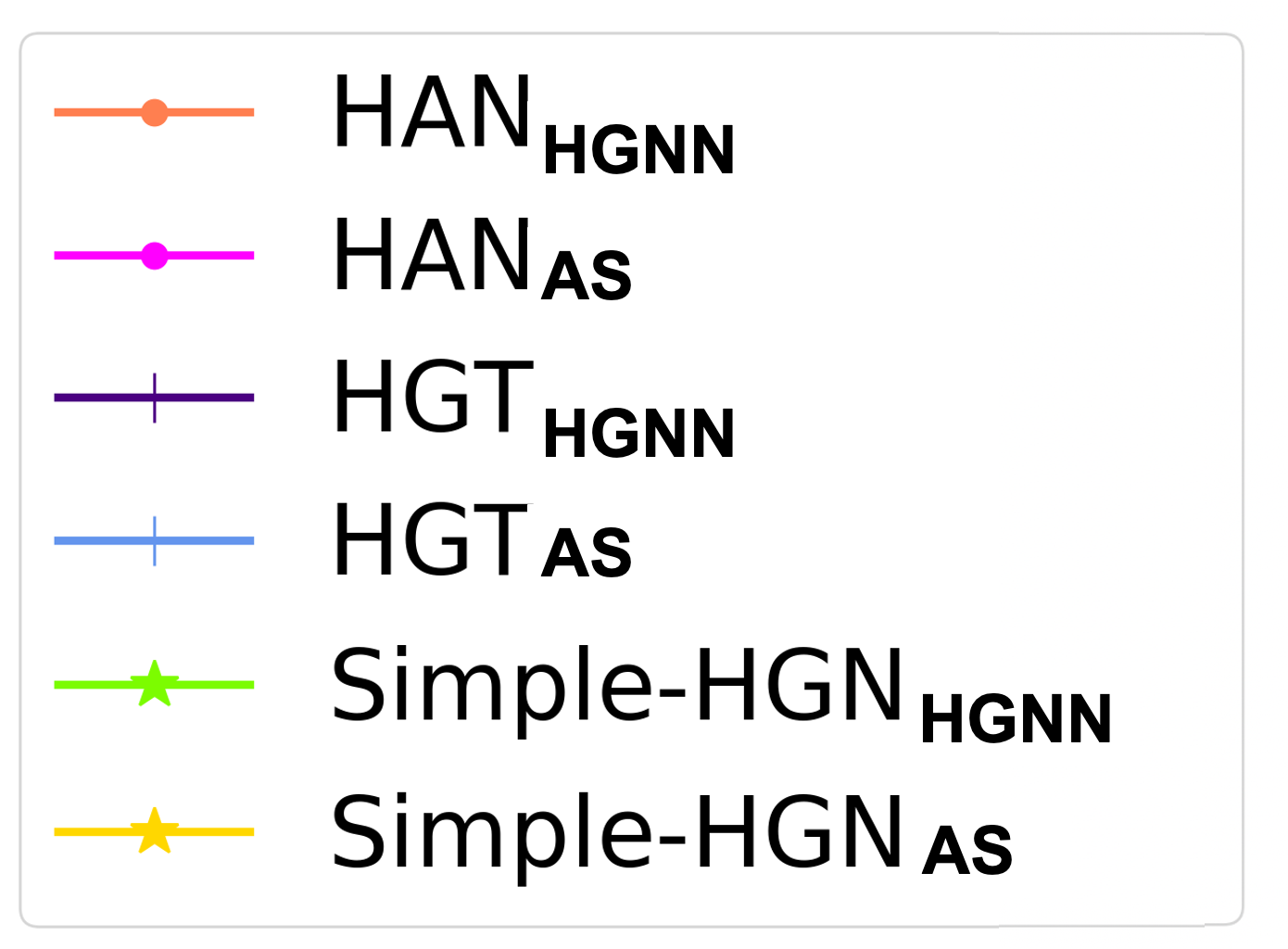}
}
\caption{Performance of \modelname{}$_\text{HGNN}$ and \modelname{}$_\text{AS}$ in $5$ iterations on ACM dataset. Iteration $0$ indicates the pretrained HGNN. Legend example: HAN$_\text{AS}$ represents the learned auxiliary system \modelname{}$_\text{AS}$ based on HAN.}
\label{Iterations}
\end{figure}

In this subsection, we will investigate the influence of two key hyper-parameters of \modelname{}

(1) \textbf{Maximum Number of Iterations in \modelname{}}:
Figure \ref{Iterations} presents the classification results of \modelname{}$_\text{HGNN}$ and \modelname{}$_\text{AS}$ in first $5$ iterations. Compared with pretrained HGNN in iteration $0$, we can see that the performance of both updated HGNN and learned auxiliary system can be effectively improved within $5$ iterations. Hence we empirically set the maximum iteration number $N=5$. 

\begin{figure}[htb]
\centering
\subfigure[20 labeled nodes per class]{
\label{20Layers}
\includegraphics[width = 0.45\columnwidth]{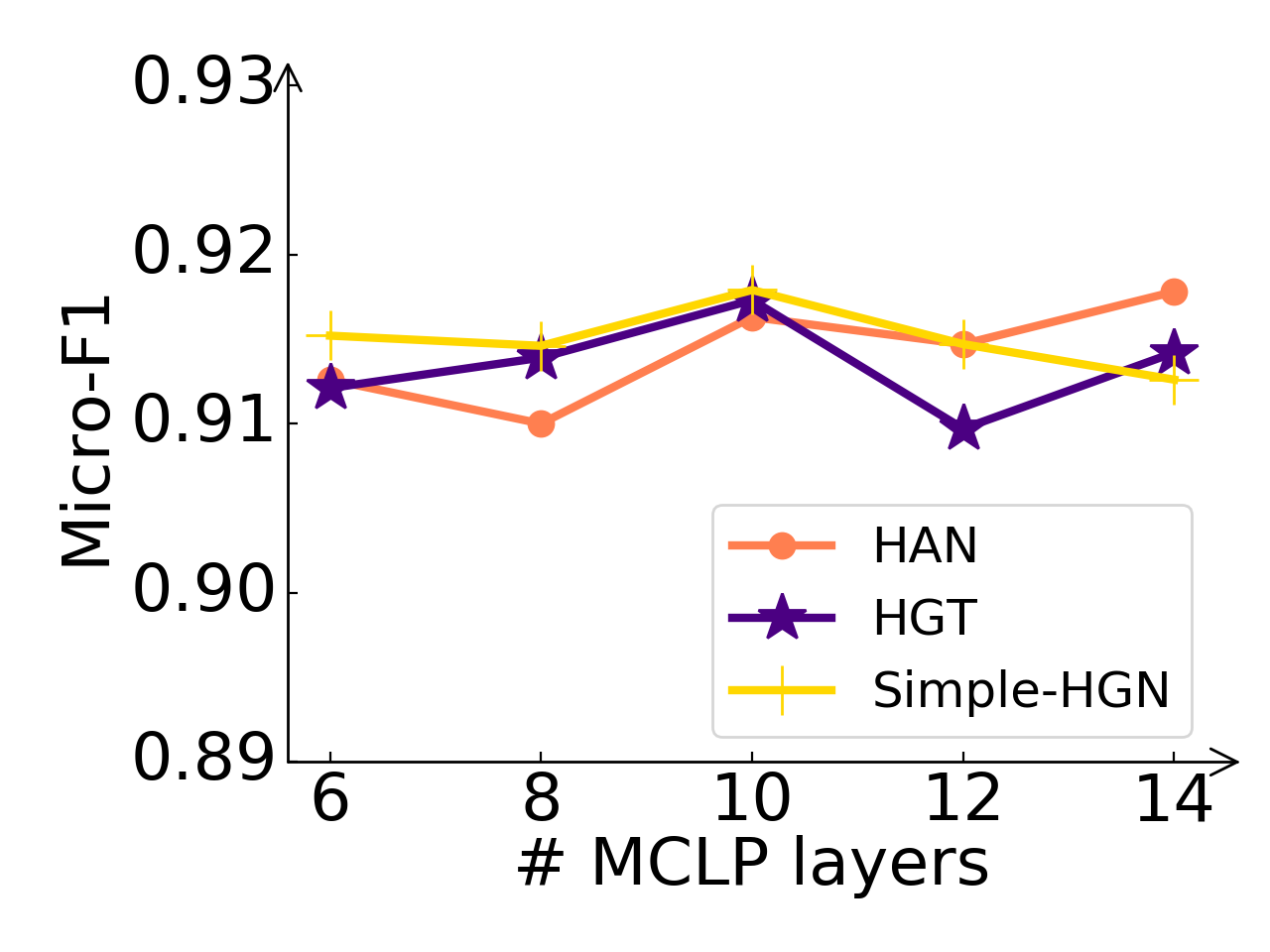}}
\hspace{-0.1cm}
\subfigure[50 labeled nodes per class]{
\label{50Layers}
\includegraphics[width = 0.45\columnwidth]{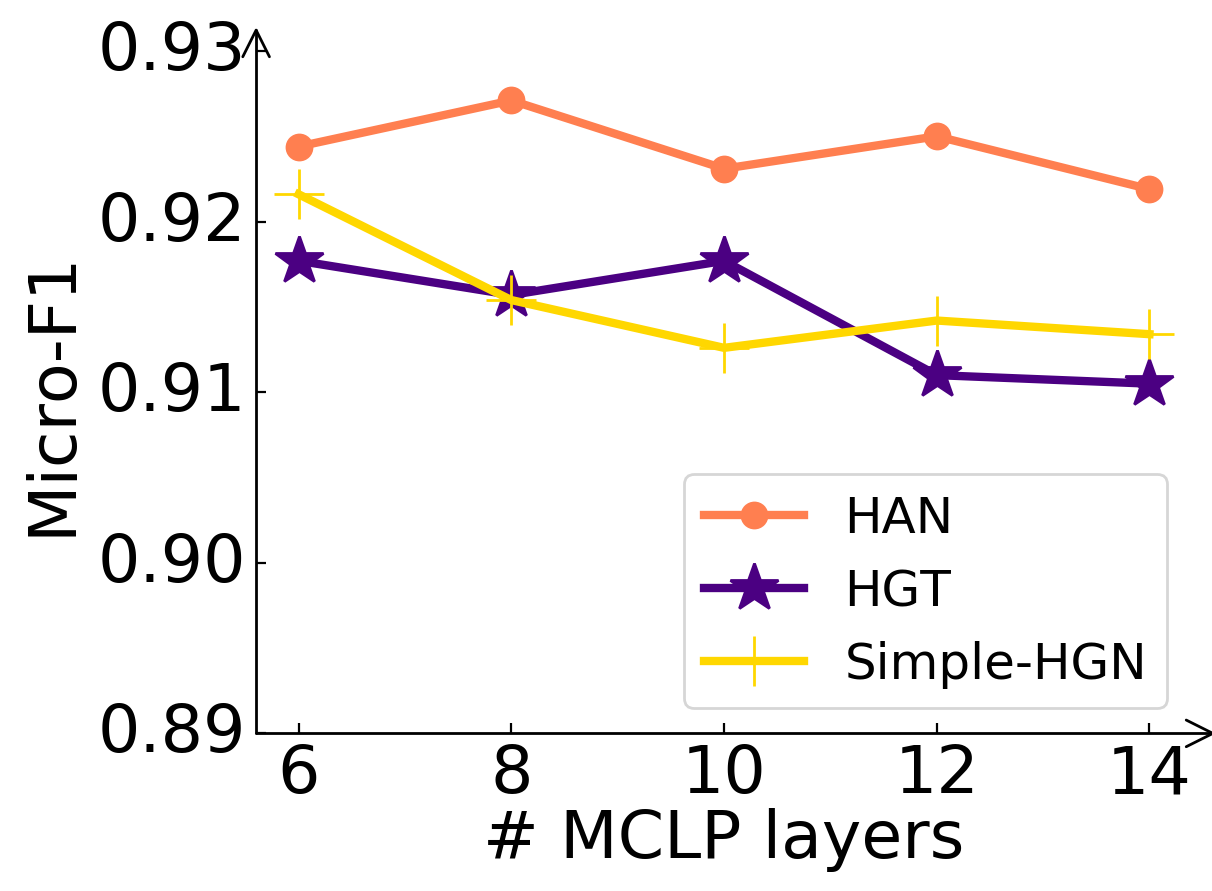}
}
\caption{Performance of \modelname{} with different layer numbers in MCLP. Legend example: HAN represents the learned auxiliary system \modelname{}$_\text{AS}$ based on HAN.}
\label{Layers}
\end{figure}

(2) \textbf{Number of Layers in MCLP}: Figure \ref{Layers} shows the performance of \modelname{} with label propagation layers $K\in\{6, 8, 10, 12,$
$14\}$. We can observe that the performance of \modelname{} is stable as the layer number $K$ changes within a reasonable range. Even the worst performance with $K$ from $6$ to $14$ has already outperformed pretrained HGNN and \modelname{}$_\text{self}$. Also, \modelname{} usually achieves the best performance with $K$ from $6$ to $10$, which is much larger than the number of stacked layers in typical HGNNs and thus enables a much wider receptive field. By propagating labels instead of representations, we barely encounter the over-smoothing issue.

\subsection{Interpretability Analysis (RQ4)}
\label{weightob}

\begin{figure}[htb]
\centering
\subfigure[Average $\gamma_v$]{
\label{Global Weights}
\includegraphics[width = 0.45\columnwidth]{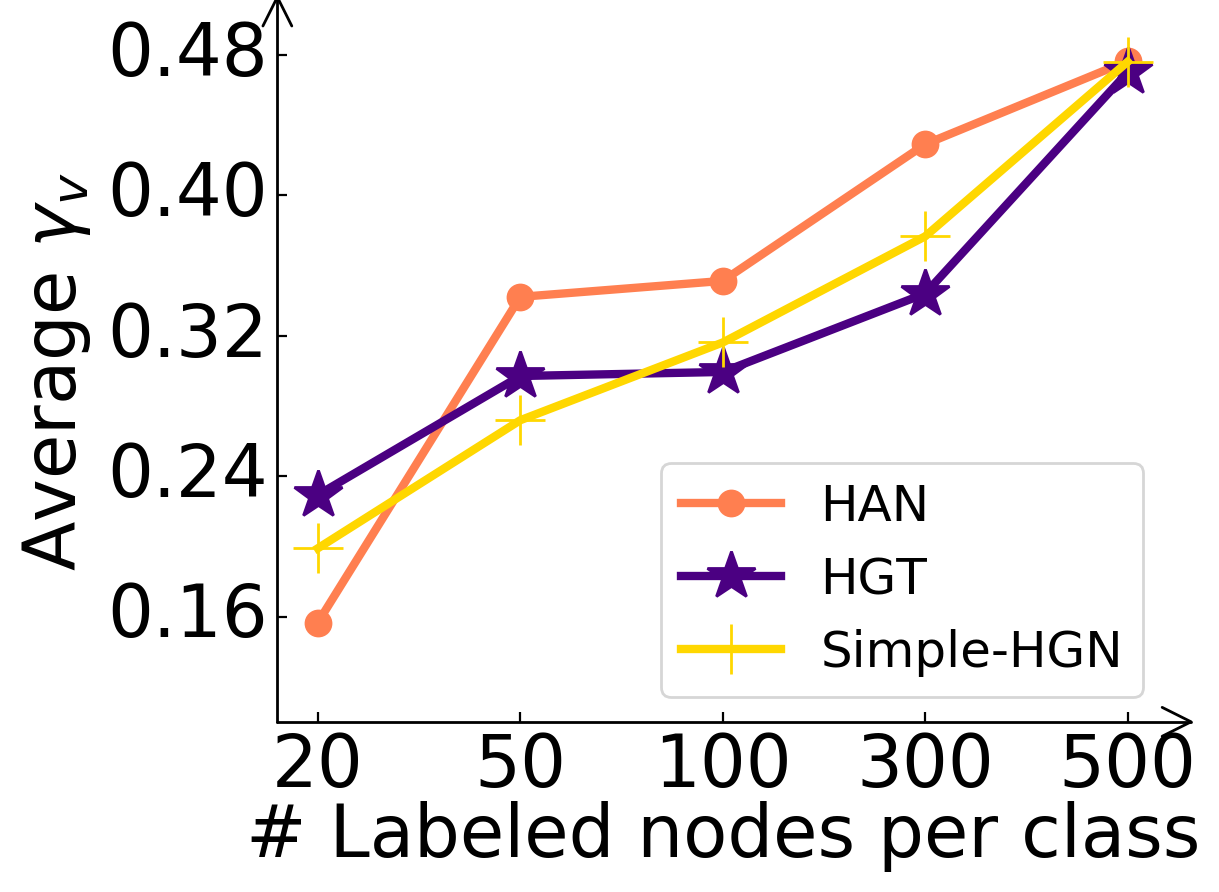}}
\hspace{-0.1cm}
\subfigure[Average $\alpha_v^P$]{
\label{MP Weights}
\includegraphics[width = 0.45\columnwidth]{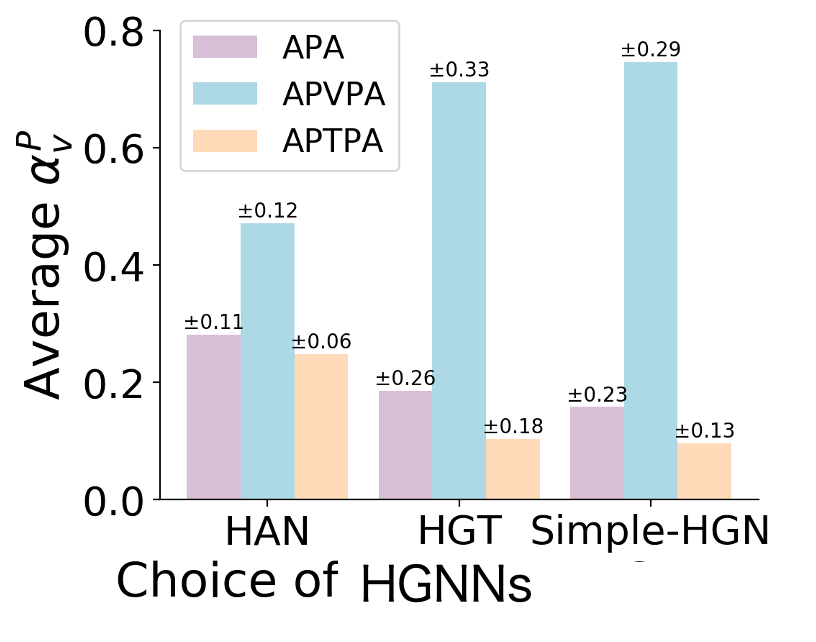}
}
\caption{Analysis of learned balance parameters, including $\gamma_v$ between global and local modules and $\alpha_v^P$ among meta-paths. We report the averages of $\gamma_v$ and $\alpha_v^P$ over all nodes. }
\label{Analysis of Weights}
\end{figure}

\noindent To answer RQ4, we conduct experiments on DBLP dataset, and investigate whether the learned auxiliary system can properly assign label recombination weights to different components. Specifically, we focus on analyzing the learned balance parameters in Eq.~\ref{Eq:Combination of Global and Local}, \ref{Eq:global prediction} and ~\ref{Eq:Local prediction}, \ie $\gamma_v$ between global and local modules, $\alpha_v^P$ among meta-paths, $\beta_v$ between a node itself and its network schema-based neighbors. We will average the balance parameters over all nodes, and report the results in Figure~\ref{Analysis of Weights}. From the results, we have the following observations: 


(1) As shown in Figure~\ref{Global Weights}, we can see that average $\gamma_v$ will increase with the number of labeled nodes, which indicates that the global module becomes more important as training ratio increases. Note that the global module will propagate labels from labeled nodes to unlabeled ones for inference. Thus, more labeled nodes will help propagate more accurately and improve the importance of the global module. In contrast, the local module is based on node features and local network schema instances, which are less sensitive to the number of labeled nodes. Therefore, our proposed \modelname{} can automatically adjust the importance of global/local modules.

(2) As shown in Figure~\ref{MP Weights}, we find that the meta-path (APVPA) always gets the largest weight $\alpha_v^P$ among the three. The partial order APVPA$>$APA$>$APTPA is consistent and independent with the choice of HGNNs. Note that the labels of authors in DBLP dataset are determined by their research areas. Hence meta-path APVPA which links two authors publishing papers in the same venue will be more suitable to propagate labels than the other two meta-paths. This observation demonstrates that our global module is able to automatically adjust the weights of different meta-paths.



(3) 
When we apply \modelname{} on HAN, HGT and Simple-HGN, the averages of $\beta_v$ are $0.4734,0.4555,0.4221$, respectively. Hence a node and its network schema-based neighbors are almost equally important in our local module, which reflects the necessity of utilizing nodes in the same network schema instances for modeling.





\section{Conclusion}
In this paper, we propose a post-training framework \modelname{} to improve HGNNs for semi-supervised node classification. To alleviate the limitations of HGNNs, we design an auxiliary system with a complementary prediction mechanism: a global inference module based on multi-channel label propagation and a local inference module based on network schema-aware prediction. Two levels of consistency can encourage the cooperation between two systems. Experimental results on three benchmark datasets with four typical HGNNs demonstrate the effectiveness of \modelname{}. For future work, an interesting direction is to generalize our framework for other graph tasks (\eg link prediction and clustering) or graph types (\eg dynamic and heterophily graphs) with proper auxiliary systems.



\begin{acks}
We would like to thank the anonymous reviewers for their valuable comments. This work is supported in part by the National Natural Science Foundation of China (No. U20B2045, 62002029, 62192784, 62172052, 62172052, U1936104).
\end{acks}

\bibliographystyle{abbrv}
\bibliography{www2023.bib}

\newpage
\;
\newpage
\appendix

\section{Pseudo Code}
 \label{code}
 Alg.~\ref{Alg_1: Cognitive System model} and ~\ref{Alg_2: Learning algorithm} show the pseudo code of our auxiliary system and the entire framework \modelname{}, respectively. 
 
  \begin{algorithm}[]
		\SetKwData{Left}{left}\SetKwData{This}{this}\SetKwData{Up}{up}
		\SetKwFunction{Union}{Union}\SetKwFunction{FindCompress}{FindCompress}
		\SetKwInOut{Input}{Input}\SetKwInOut{Output}{Output}
		
		\Input{
			HIN $\mathcal{G} = \{\mathcal{V},\mathcal{E},\mathcal{T},\mathcal{R},\phi,\varphi\}$,\\
			node features ${\mathcal{X}}={\{\mathbf{x}_v|\forall{v\in\mathcal{V}}\}}$,\\
			target node type ${T \in {\mathcal{T}}}$,\\
			labeled node set ${\mathcal{V}}_{L}\subset{{\mathcal{V}}_{T}}$,\\ 
			unlabeled node set ${\mathcal{V}}_{U}={{\mathcal{V}}_{T}}\setminus \mathcal{V}_{L}$,\\ 
			metapath set ${\mathcal{P}}$,\\
			number of layers $K$ in the global module;\\}
		\Output{Predicted label distribution $g_{\Omega}(v)$ for every unlabeled node $v\in\mathcal{V}_{U}$;}
		\tcp{Global Inference Module: }
		\For{$P\in{\mathcal{P}}$}{
			Initialize the label prediction $l^{0}_{P}(v)$ for $v\in{\mathcal{V}_{T}}$ by Eq.~\ref{Eq:label initialize(LP)};\\
			
			\For{$k=1,2\dots K$}{
				\For{$v\in{\mathcal{V}_{U}}$}{
					Update the label prediction of $v$ by Eq.~\ref{Eq:LP function(k->k+1)};\\
				}
			}
		}
		\For{$v\in{\mathcal{V}_{U}}$}{
			Combine the predictions from different channels as the global-level output by Eq.~\ref{Eq:global prediction};\\
		}
		\tcp{Local Inference Module: }
		\For{$v \in \mathcal{V}$}{
			Perform type-specific feature transformation by Eq.~\ref{Eq:feature projection};\\
			Perform label projection by Eq. \ref{Eq:label_projection};\\
		}
		\For{$v\in{\mathcal{V}_{U}}$}{
			Compute the local-level output by Eq. \ref{Eq:Local prediction}:\\
		}
		\tcp{Combination of Two Modules: }
		\For{$v\in{\mathcal{V}_{U}}$}{
			Compute the prediction of auxiliary system by Eq.~\ref{Eq:Combination of Global and Local};\\
		}
        \caption{The Implementation of auxiliary system}
        \label{Alg_1: Cognitive System model}
 \end{algorithm}

 \begin{algorithm}[]
		\SetKwData{Left}{left}\SetKwData{This}{this}\SetKwData{Up}{up}
		\SetKwFunction{Union}{Union}\SetKwFunction{FindCompress}{FindCompress}
		\SetKwInOut{Input}{Input}\SetKwInOut{Output}{Output}
		
		\Input{
			HIN $\mathcal{G} = \{\mathcal{V},\mathcal{E},\mathcal{T},\mathcal{R},\phi,\varphi\}$,\\
			node features ${\mathcal{X}}$,\\
			labeled node set ${\mathcal{V}}_{L}\subset{{\mathcal{V}_T}}$,\\
			unlabeled node set ${\mathcal{V}}_{U}=\mathcal{V}_T\setminus \mathcal{V}_{L}$,\\
			validation node set ${\mathcal{V}}_{D}\subset{{\mathcal{V}_U}}$,\\
			test node set ${\mathcal{V}}_{S}=\mathcal{V}_U\setminus \mathcal{V}_{D}$,\\
			number of iterations $N$, number of epochs $M$;\\
		}
		\Output{Trained node label predictors $f_{\Theta},g_\Omega$.  }
		Pretrain the HGNNs with labeled node set $\mathcal{V}_{L}$ by Eq.~\ref{Eq:pre-train obejective}\;
		\For {$i=1,2\dots N$}{
		\For {$epoch=1,2\dots M$}{
			Update parameter $\Omega$ of auxiliary system by Eq.~\ref{Eq:Cognitive System objective};\\
			Evaluate $\Omega$ on validation set $\mathcal{V}_{D}$;\\
			}
			Select $\Omega$ as the epoch with the best performance on $\mathcal{V}_{D}$;\\
			\For {$epoch=1,2\dots M$}{
			Update parameter $\Theta$ of HGNNs by  Eq.~\ref{Eq:Perceptive System objective};\\
			Evaluate $\Theta$ on validation set $\mathcal{V}_{D}$;\\
			}
			Select $\Theta$ as the epoch with the best performance on $\mathcal{V}_{D}$;\\
		}
		\caption{The Overall Framework of \modelname{}}
		\label{Alg_2: Learning algorithm}
\end{algorithm}

\section{Details and Statistics of Datasets}
\label{datasets}
Table~\ref{Tab:Dataset statistic whole} shows the statistics of three datasets, and the detailed descriptions about the three datasets are as follows:
\begin{table*}[]
  \centering
  \caption{Dataset statistics with more details.}
    \begin{tabular}{c|c|c|c|c|c|c|c|c}
    \toprule[1pt]
    \textbf{Dataset} & \textbf{\# Nodes} & \textbf{\# Edges} & \textbf{\# Features} & \textbf{\# Meta-paths} & \textbf{\# Classes} & \textbf{\# Training} & \textbf{\# Validation} & \textbf{\# Test} \\
    \midrule
    \textbf{ACM} & \tabincell{c}{Author (7167)\\
Paper (4019)\\
Subject (60)} & \tabincell{c}{P-A (13407)\\
P-S (4019)} & 1902  & \tabincell{c}{P-A-P\\
P-S-P} & 3     & \tabincell{c}{20 $\times$ 3\\
50 $\times$ 3} & \tabincell{c}{50 $\times$ 3\\
50 $\times$ 3} & \tabincell{c}{3809\\
3719} \\
    \midrule
    \textbf{DBLP} & \tabincell{c}{Author (4057)\\
Paper (14328)\\
Term (7723)\\
Venue (20)} & \tabincell{c}{P-A (19645)\\
P-T (85810)\\
P-V (14328)} & 334   & \tabincell{c}{A-P-A\\
A-P-V-P-A\\
A-P-T-P-A} & 4     & \tabincell{c}{20 $\times$ 4\\
50 $\times$ 4} & \tabincell{c}{50 $\times$ 4\\
50 $\times$ 4} & \tabincell{c}{3777\\
3657} \\
    \midrule
    \textbf{IMDB} & \tabincell{c}{Movie (4278)\\
Director (2081)\\
Actor (5257)} & \tabincell{c}{M-D (4278)\\
M-A (12828)} & 3066  & \tabincell{c}{M-A-M\\
M-D-M} & 3     & \tabincell{c}{20 $\times$ 3\\
50 $\times$ 3} & \tabincell{c}{50 $\times$ 3\\
50 $\times$ 3} & \tabincell{c}{4068\\
3978} \\
    \bottomrule[1pt]
    \end{tabular}%
  \label{Tab:Dataset statistic whole}%
\end{table*}%

 $\bullet$ \textbf{ACM}\footnote{https://github.com/Andy-Border/NSHE}~\cite{NSPHINE}
 is a citation network with the target node type as Paper (P). All the papers in the HIN are divided into three classes: database, wireless communication, and data mining. We use the bag-of-words representations of keywords as the node features of paper nodes. 
 
 $\bullet$ \textbf{DBLP}\footnote{https://github.com/cynricfu/MAGNN}~\cite{MAGNN}
 is a bibliography website of computer science. The target node type is Author (A), and the authors are divided into four classes according to their research areas (database, data mining, artificial intelligence, and information retrieval). The features of authors are also described by bag-of-words representations of their paper keywords. 
 
 $\bullet$ \textbf{IMDB}\footnote{https://github.com/cynricfu/MAGNN}~\cite{MAGNN}
 is a website about movies and television programs. We use an HIN with the target node type as Movie (M). All the movies are labeled as one of three classes (action, comedy, and drama) according to their genres. The features of movies are described by a bag-of-words representation of its plot keywords.
 
 For all three datasets, the features of nodes with other types are one-hot vectors. 
 
\section{Implementation details}
\label{implemention details}
We implement \modelname{} based on PyTorch and Deep Graph Library (DGL)\footnote{https://github.com/dmlc/dgl}. For all experiments, we employ the GPU device of GeForce RTX 2080 and 3090.

\section{Details of Experimental Settings}

\label{other setting}
In this section, we will describe detailed settings of our \modelname{}. 

\subsection{HGNNs Settings}
In this subsection, we provide the detailed settings of the four HGNNs used in our \modelname{}.

$\bullet$ \textbf{HAN}~\cite{HAN} is an HGNN model which learns meta-path-specific node representations by leveraging node-level attention and semantic-level attention mechanism. We use $8$ attention heads and 64-dimensional hidden size in our experiments. 

$\bullet$ \textbf{HGT}~\cite{HGT} is a transformer-based HGNN model with heterogeneous subgraph sampling. We employ a 2-layer HGT with $4$ to $8$ attention heads and 256-dimensional hidden size in our experiments. 

$\bullet$ \textbf{Simple-HGN}~\cite{SimpleHGN} is an improved version of GAT~\cite{GAT} with learnable edge-type embedding and residual connections. In our experiments, we use a 2-layer Simple-HGN with attention heads from $4$ to $8$, hidden size from $\{64, 128, 256\}$ and 32-dimensional edge-type embedding.

$\bullet$ \textbf{MAGNN}~\cite{MAGNN} is an improved HAN with several meta-path encoders to embed all the nodes along a meta-path. In our experiments, we use $8$ attention heads, hidden size from $\{64, 128, 256\}$, attention vector dimension from $\{32, 64, 128\}$, and batch size from $\{16, 32, 64, 128,$ $ 256,512\}$.

For the pretraining settings of HGNNs, we will run HGNN models for $150$ epochs, and choose the best epoch according to the performance on the validation set. For hyper-parameter settings, we use dropout rate from $0.2$ to $0.8$, learning rate from $0.001$ to $0.05$, and weight decay of Adam optimizer from $\{0,0.0001,0.0005,0.001\}$.

We also tried GTN~\cite{GTN}, but omit its results since GTN performs worse (either with or without \modelname{}) than the above four HGNNs under our experimental settings.
\subsection{Auxiliary System Settings}
We explore dropout rate of local module from $\{0.4, 0.5, 0.6, 0.7\}$, and use Adam optimizer with learning rate as 0.01 for training, the weight decay of the optimizer is set as 0.0005 for ACM/IMDB and 0 for DBLP. 

\subsection{Optimization Settings} For distance functions, we use Euclidean distance between two systems, and KL-divergence between global and local modules in Eq. \ref{Eq:Cognitive System objective}. For Eq. \ref{Eq:Perceptive System objective}, we employ the KL-divergence function. We use Xavier normal distribution to initialize parameters. 

\subsection{Detailed Settings in Section~\ref{SOTA-exp}}
\label{SOTA_setting}
In this subsection, we provide the detailed settings of the models in Section~\ref{SOTA-exp}.

$\bullet$ \textbf{GCN}~\cite{GCN}: We set hidden size $k=128$ for all datasets. For ACM, we set layers $L=2$,  we set layers $L=3$ for DBLP and IMDB.

$\bullet$ \textbf{GAT}~\cite{GAT}: We set hidden size $k=64$ , layers $L=3$, the number of attention heads $n=4$ and negative slope $s=0.05$ for all datasets.  

$\bullet$ \textbf{CPF}~\cite{yang2021extract}: We set hidden size $k=64$, LP layers $L=5$, MLP layers $l=2$ for all datasets.

$\bullet$ \textbf{RDD}~\cite{zhang2020reliable}: We set hidden size $k=64$, parameters $\gamma=3$ and $\beta=10$ for all datasets.

$\bullet$ \textbf{HGNN-AC}~\cite{HGNN-AC}: We employ MAGNN as the backbone of HGNN-AC, and we set hidden size $k=128$, attention vector size $a=128$, and the number of attention heads $n=8$ for all datasets. For DBLP, we set the batch size $bz=8$. 

$\bullet$ \textbf{HeCo}~\cite{HeCo}: We set hidden size  $k=256$ for ACM and IMDB, for DBLP, we set $k=64$.

$\bullet$ \textbf{HGSL}~\cite{HGSL}: We set hidden size  $k=64$ for DBLP, for ACM and IMDB, we set $k=256$. We set the number of heads $h=2$ for all datasets.

\subsection{Detailed Settings in Section~\ref{weightob}}
For $\gamma_v$ between global and local modules in Eq.~\ref{Eq:Combination of Global and Local}: We conduct experiments by gradually increasing the number of labeled nodes, and report the changing trend of the average of $\gamma_v$ over all node $v\in{\mathcal{V}_T}$. 

For $\alpha_v^P$ among meta-paths in Eq.~\ref{Eq:global prediction}: To prove that our global inference module is capable of identifying the importance of different meta-paths, we report the average of $\alpha_v^P$ over all node $v\in{\mathcal{V}_T}$ for every meta-path $P$. Here we use 100 labeled nodes per class for training, because there are larger weights of the global module (\ie $\gamma_v$) under this setting and the difference of $\alpha_v^P$ will contribute more to the final prediction of auxiliary system. In addition, we also mark the standard deviation of $\alpha_v^P$ on each column in Figure~\ref{MP Weights}. We can see that the weights of different nodes can vary greatly, which means that it is necessary to use node-specific balance parameters for modeling. 

For $\beta_v$ which balances the importance of the label distribution of a node itself and its network schema-based neighbors in Eq.~\ref{Eq:Local prediction}: We also use 100 labeled nodes per class as $\alpha_v^P$ for training for consistency.

\section{Time and memory cost of HGPF}

We report the time cost to run a single epoch for the four HGNNs and AS on the ACM dataset in Table \ref{tab_time}. The running speed of AS is second only to SimpleHGN and much faster than that of MAGNN and HGT. In addition, we also compared the memory consumption of HGPF and pretrained HGNNs on the ACM dataset in Table \ref{tab_memory}. The memory cost of HGPF is similar to that of pretrained HGNNs, which indicates that the memory overhead of HGPF is negligible.

\begin{table}[htbp]
  \centering
  \small
  \caption{Training time per epoch.}
    \begin{tabular}{c|c|c|c|c|c}
    \toprule[1pt]
    \textbf{} & \textbf{HAN} & \textbf{MAGNN} & \textbf{SimpleHGN} & \textbf{HGT} & \textbf{AS} \\
    \midrule
    \textbf{Time} & 0.185s & 0.581s  & 0.093s
 & 0.331s & 0.171s \\
    \bottomrule[1pt]
    \end{tabular}
  \label{tab_time}%
\end{table}

\begin{table}[htbp]
  \centering
  \small
  \caption{Memory cost during training.}
    \begin{tabular}{c|c|c|c|c}
    \toprule[1pt]
    \textbf{} & \textbf{HAN} & \textbf{MAGNN} & \textbf{SimpleHGN} & \textbf{HGT} \\
    \midrule
    \textbf{Pretrain} & 4591MB & 9108MB  & 3865MB
 & 5791MB \\
    \midrule
    \textbf{HGPF} & 4665MB & 9243MB  & 3873MB
 & 5846MB \\
    \bottomrule[1pt]
    \end{tabular}
  \label{tab_memory}%
\end{table}

\section{Case study}

As illustrated in the introduction section, there are two types of nodes difficult to predict by HGNNs: (1)  nodes far from training nodes of same category; (2) nodes with more training nodes from different categories than the same category in the receptive field. Here we present case study for such ``hard nodes'' to show the effectiveness of HGPF.
We select two representative nodes from each type on ACM dataset for analysis: HAN predicts these cases incorrectly, while HGPF can predict their accurate labels. 

\textbf{Case 1 (ID 2883)}: The node has a distance greater than 4 from all training nodes of the same category. In HGPF, the node-specific balance parameter $\gamma_v = 0.59$, which means that the prediction of this case is more dependent on the global module to capture more information from remote training nodes. And the meta-path balance parameter $\alpha^{PAP}_v = 0.41$, $\alpha^{PSP}_v = 0.59$, which means meta-path PSP is more helpful for this case. 

\textbf{Case 2 (ID 2948)}: Similar to case 1, there is no training node with a distance less than or equal to 4 from this node. The parameter $\gamma_{v}$ of this case is 0.73, the prediction of global module also has larger weight than local module. And the meta-path balance parameter $\alpha^{PAP}_v = 0.49$, $\alpha^{PSP}_v = 0.51$, which means that both meta-paths PAP and PSP are important for this node. 

\textbf{Case 3 (ID 3499)}: Among the 4-hop neighbors of this case, the numbers of training nodes with label 0, 1, and 2 are 4, 20, and 3 respectively, and HAN incorrectly predicts it as label 1, since the training node neighbors with label 1 disturb the prediction of HGNNs. For HGPF, the $\gamma_{v}$ of this case is 0.31, which means that the local module play a more important role in the prediction of this case than global module. The network schema based local module can effectively reduce the noise cause by training node neighbors from different categories. 

\textbf{Case 4 (ID 3032)}: Similar to case 3, in its 4-hop neighborhood, the number of training nodes with label 0, 1, and 2 are 15, 1, and 12 respectively.  In this case, the $\gamma_{v}$ of this case is only 0.13, the prediction highly relies on the local module to reduce the noise from training node neighbors from different categories. 

As shown from the above case studies, for HGPF, the prediction of case 1 and case 2 (type 1) is more dependent on the global inference module, while the prediction of case 3 and case 4 (type 2) is more dependent on the local inference module, which is consistent with our motivation.

\end{document}